\documentclass[preprint,prd,aps,showpacs,preprintnumbers,amsmath,amssymb]{revtex4-1}
\usepackage{graphics,epsfig,subfigure}
\usepackage{diagbox}
\usepackage[usenames]{color}
\usepackage{color}
\usepackage{graphicx}
\usepackage{amsfonts}
\usepackage{indentfirst}
\usepackage{booktabs}
\usepackage{hyperref}
\hypersetup{hypertex=ture,backref=true,colorlinks=ture,linkcolor=blue,anchorcolor=blue,citecolor=blue}
\usepackage{float}
\usepackage{multirow}
\usepackage[section]{placeins}
\usepackage{array}
\usepackage{amsmath}

\begin{document}
	\title{\Large \bf Axion boson stars with wormhole topology}
	\author{Chen-Hao Hao$^{1,4}$, Yong-Qiang Wang$^{2,3}$, Jieci Wang$^{1,4 *}$}
	\email{jcwang@hunnu.edu.cn, corresponding author}
	\affiliation{$^{1}$Department of Physics, Key Laboratory of Low Dimensional Quantum Structures and Quantum  Control of Ministry of Education, and Synergetic Innovation Center for Quantum Effects and Applications, Hunan Normal
		University, Changsha, Hunan 410081, P. R. China\\
		$^{2}$Lanzhou Center for Theoretical Physics, Key Laboratory of Theoretical Physics of Gansu Province,
		School of Physical Science and Technology, Lanzhou University, Lanzhou 730000, China\\
		$^{3}$Institute of Theoretical Physics $\&$ Research Center of Gravitation, Lanzhou University, Lanzhou 730000, China\\
			$^{4}$ Institute of Interdisciplinary Studies, Hunan Normal University, Changsha, Hunan 410081, P. R. China}
	
	\begin{abstract}
		We investigate a novel gravitational configuration formed by a massless real phantom field and an axion scalar field, minimally coupled to gravity. This system describes an Ellis-type wormhole situated at the center of an axion star. By normalizing the mass of the axion field to unity, the physical properties of the model are determined by three independent parameters: the potential's decay constant, the frequency of the axion field, and the wormhole's throat parameter. We assess the traversability of this wormhole by examining the curvature scalars and energy conditions of the static solution. Our analysis of the wormhole's embedding diagrams indicates that, although the wormhole typically exhibits a single-throat geometry, a double-throat configuration featuring an equatorial plane may arise under specific conditions. Finally, an analysis of the null-geodesics reveals the existence of at least one unstable light ring at the wormhole throat.
	\end{abstract}

	\maketitle

	\section{INTRODUCTION}\label{Sec1}
	
	The twin puzzles of dark matter and dark energy stand as formidable challenges at the frontiers of modern cosmology and gravitational theory. Observations of Type Ia supernovae (SNeIa) \cite{SupernovaSearchTeam:2001qse,Perlmutter:1999jt} and independent evidence from the cosmic microwave background (CMB) \cite{WMAP:2003ivt,WMAP:2003zzr} have provided compelling support for the existence of a negative-pressure component known as ``dark energy." This dark energy can be simply parameterized by its equation of state, $k=p_d/\rho_d$, where $p_d$ is  the spatially homogeneous
	pressure and $\rho_d$ is the dark energy density \cite{Cai:2004dk}. For cosmic expansion, the condition $k<-1/3$ must be satisfied, with $k=-1$ corresponding to the cosmological constant \cite{Sahni:2002kh}. In addition to the standard candidate range of $-1 \le k <-1/3$ \cite{Copeland:2006wr,Frieman:2008sn}, recent theoretical considerations have also explored the case where $k<-1$ \cite{Melchiorri:2002ux,Alcaniz:2003qy,Carroll:2003st}. This latter scenario violates the null energy condition and other energy conditions. This hypothetical substance is often termed ``phantom energy" and could mediate a long-range repulsive force \cite{Amendola:2004qb}, a possibility that could not be observationally ruled out even in recent measurements \cite{Planck:2015fie}.
	
	While cosmological phantom energy models face significant theoretical challenges, the underlying mechanism—a scalar field with a negative kinetic term—provides a straightforward and widely used theoretical tool for modeling the ``exotic matter" required for the construction of traversable wormholes \cite{Ellis:1973yv,Ellis:1979bh,Kodama:1978dw}. The wormhole concept originates from the work of Einstein \cite{Einstein:1935tc}, with the term coined by Wheeler \cite{Misner:1957mt}, and is of significant interest in various models of quantum gravity \cite{Visser:1995cc}. For the traversable wormholes, the simplest scenario involves a phantom scalar field acting as the source of such exotic matter, thereby providing the necessary support for the wormhole structure \cite{Lobo:2005us,Lobo:2005yv,Yazadjiev:2017twg}. Following the seminal work of  Morris and Thorne in 1988 \cite{Morris:1988cz}, research on traversable wormholes has expanded to include numerous aspects \cite{Bronnikov:2012ch,Kleihaus:2014dla,Novikov:2009vn,Bronnikov:2013coa,Huang:2020qmn,Bronnikov:2002rn,Kanti:2011jz,Blazquez-Salcedo:2020nsa,DeFalco:2023twb,DeFalco:2023kqy}. More recently, various models for traversable wormholes have been proposed that either do not violate the energy condition \cite{KordZangeneh:2015dks} or do not require exotic matter \cite{Harko:2013yb,Konoplya:2021hsm,Blazquez-Salcedo:2020czn,Kain:2023pvp}. 
	
	Parallel to the mystery of dark energy is the enduring puzzle of dark matter. Among the myriad proposed candidates, scalar field models have gained significant traction \cite{Hu:2000ke,Sahni:1999qe,Matos:2000ng}. Some of these scalar fields can form extended, gravitationally bound objects, ranging in size from microscopic particles to vast galactic halos. These are known as boson stars \cite{Kaup:1968zz,Ruffini:1969qy,Schunck:1996he,Liebling:2012fv}, and these extended compact objects have the potential to mimic the observational signatures of dark matter \cite{Davidson:2016uok}. However, one of the most compelling current dark matter candidates is the axion \cite{Baer:2014eja,Klaer:2017ond}. The axion is a pseudo-Nambu-Goldstone boson whose non-derivative coupling to the Standard Model arises solely from topological charges \cite{Preskill:1982cy}. While originally proposed by Peccei and Quinn to resolve the strong CP problem in Quantum Chromodynamics (QCD), axions have since become a paradigm for ultralight, feebly interacting bosons beyond the Standard Model \cite{Jaeckel:2010ni,Arvanitaki:2010sy,Berezhiani:1992rk,Sakharov:1996xg,Khlopov:1999tm}.
	
	Analogous to other scalar theories, the axion field can form its own gravitationally bound configurations known as axion stars  \cite{Visinelli:2017ooc,Braaten:2015eeu,Eby:2016cnq,Guerra:2019srj}, arising from the minimal coupling of the axion field to gravity. Recent work has investigated rotating axion stars \cite{Delgado:2020udb}, as well as multi-field configurations that involve the mixing of a rotating axion star with other bosonic fields and the tidal effects on axion stars \cite{Zeng:2021oez,Chen:2023vet}.
	A compelling avenue of research opens when we move beyond studying these phenomena in isolation and instead consider hybrid configurations: nontrivial spacetime, such as wormholes supported by phantom fields, that also harbor additional matter fields. Although these components are traditionally considered on different physical scales—from the cosmological to the astrophysical, studying their interplay in a single, localized system serves as a valuable theoretical laboratory. Previous work has already explored wormhole systems coupled to ordinary scalar fields \cite{Dzhunushaliev:2014bya,Ding:2023syj,Hao:2024hba}, Proca fields \cite{Su:2023zhh}, fermionic fields \cite{Hao:2023igi}, and nonlinear electromagnetic fields \cite{Su:2024gxp}. These studies underscore a key insight: introducing a wormhole almost invariably alters the properties of the corresponding gravitational solution in trivial topology \cite{Hoffmann:2017jfs}, and the specific nature of the coupled matter field has a significant impact on the resulting wormhole spacetime \cite{Dzhunushaliev:2025fbf,Hao:2023kvf}.
	
	Motivated by the possibility of non-trivial interactions within the dark sector \cite{Wang:2016lxa}, it is compelling to investigate models where candidates for these phenomena are coupled in extreme gravitational environments. Therefore, in this paper, we numerically construct and analyze a spherically symmetric, asymptotically flat configuration composed of an axion scalar field and a massless real phantom field minimally coupled to gravity. We systematically investigate how three key parameters: the decay constant $f_a$, the field frequency $\omega$, and the throat parameter $r_0$ influence the system's physical properties. The traversability of the wormhole is evaluated through an analysis of the curvature scalar and energy conditions of the static solution. Finally, we study the model's null-geodesics and the corresponding light rings.
	
	The paper is organized as follows. In Sec.~\ref{sec2}, we present the model four-dimensional
	Einstein gravity coupled to a phantom field and a axion field. In Sec.~\ref{sec3}, the boundary conditions are studied. The numerical results of the three different cases are discussed in Sec.~\ref{sec4}. We conclude in Sec.~\ref{sec5} with a summary and illustrate the range for future work.

	\section{THE MODEL}\label{sec2}

	\subsection{Action}

	We consider the Einstein-Hilbert action including the Lagrangian for the axion field and the phantom scalar field, the action is given by
	\begin{equation}\label{action}
		S=\int\sqrt{-g}d^4x\left(\frac{R}{2\kappa}+\mathcal{L}_{p}+\mathcal{L}_{a}\right),
	\end{equation}
	where $R$ is the Ricci scalar.  The term
	$\mathcal{L}_{p}$ and $\mathcal{L}_{a}$ are the Lagrangians defined by
	
	\begin{eqnarray}
		\nonumber	\mathcal{L}_{a}  &= & -\nabla_a\Psi^*\nabla^a\Psi  - V(|\Psi|^2),  \\
		\mathcal{L}_{p}  & =   &   \nabla_a\Phi\nabla^a\Phi.
	\end{eqnarray}
	
	Here $\Psi$  and $\Phi$ represent the complex axion scalar field and the phantom field, respectively.
	By varying the action (\ref{action}) with respect to the metric, we can obtain the Einstein equations
	\begin{equation}
		\label{eq:EKG1}
		R_{\mu\nu}-\frac{1}{2}g_{\mu\nu}R-\kappa T_{\mu\nu}=0,
	\end{equation}
	with stress-energy tensor
	\begin{equation}
		T_{\mu\nu} = g_{\mu\nu}({{\cal L}}_a+{{\cal L}}_p)
		-2 \frac{\partial ({{\cal L}}_a+{{\cal L}}_p)}{\partial g^{\mu\nu}}.
	\end{equation}
	
	The equations governing the matter fields are derived by performing variations with respect to both the phantom field and the axion field, which are
	\begin{equation}
		\label{eq:EKG2}
		\Box\Psi-\dfrac{\partial V}{\partial |\Psi|^2} \Psi = 0,
	\end{equation}
	and
	\begin{equation}
		\label{eq:EKG3}
		\Box\Phi=0.
	\end{equation}
	
	\subsection{Ansatze}
	
	We consider the  general static spherically symmetric solution with a wormhole,
	and adopt the Ansatzes as follows ~\cite{Hoffmann:2017jfs}
	\begin{equation}  \label{line_element1}
		ds^2 = -e^{A}  dt^2 +C e^{-A}   \left[ d r^2 + h (d \theta^2+\sin^2 \theta d\varphi^2) \right]\,,
	\end{equation}
	where $A$ and $C$ are functions of  radial coordinate $r$,  $h=r^2+r_0^2$ with  the throat parameter  $r_0$,
	and $r$  ranges from positive infinity to negative infinity.
	It should be emphasized that the two limits $r\rightarrow \pm\infty$ correspond to two distinct asymptotically flat spacetime.
	
	Furthermore, we assume stationary axion complex scalar  field and phantom field in the form
	\begin{eqnarray}  \label{an2}
		\Psi&=\psi(r)e^{i\omega t}, \;\;\;\;  \Phi&=\phi(r).
	\end{eqnarray}
	Here, $\psi$ is only a real function of the radial coordinate $r$, and the constant $\omega$ is referred to as the synchronization frequency. Moreover, the phantom field $\Phi$ is also a real function and is independent of the time coordinate $t$. The potential of axion field is 
	
	\begin{equation}
		V(\psi) = \frac{2 \mu^2 f_a^2}{B} \left[ 1 - \sqrt{1 - 4 B \sin^2 \left( \frac{\psi}{2 f_a} \right)} \right],
		\label{Potential}
	\end{equation}
	where $B$ is a constant related to the ratio of the up quark mass $m_u$ to the down quark mass $m_d$, with $m_u/m_d \approx 0.48$, giving $B \approx 0.22$, $\mu$ and $f_a$ are two free parameters. In this potential, the second term corresponds to the QCD axion effective potential~\cite{GrillidiCortona:2015jxo}, and the addition of a constant term ensures $V(0)=0$, in order to construct asymptotically flat axion stars. Expanding the potential around $ \psi_0=0 $, one obtains
	\begin{equation}
		V(\psi) = \mu^2 \psi^2 - \left( \frac{3B-1}{12} \right) \frac{ \mu^2}{f_a^2} \psi^4 + \dots \ .
		\label{UnfoldingPotential}
	\end{equation}
	It can be observed that $\mu$ represents the mass of the axion, while $f_a$ denotes the decay constant of the axion field. When $f_a \gg \psi$, only the free scalar potential remains, and the model reduces to the nontrivial topology mini-boson stars~\cite{Dzhunushaliev:2014bya,Ding:2023syj,Hao:2024hba}.

	\subsection{Equations}
	
	Substituting the Eq.(\ref{an2}) and Eq.(\ref{line_element1}) into the Eq.(\ref{eq:EKG2}) and Eq.(\ref{eq:EKG3}) leads to 
	\begin{equation}
		\label{mat}
		- \frac{f_a \mu^2 \sin\left(\frac{\psi}{f_a}\right)}{\sqrt{1 - 2B + 2B \cos\left(\frac{\psi}{f_a}\right)}} + e^{-A} \omega^2 \left[ \left( \frac{2e^A r}{hC} + \frac{e^A C'}{2C^2} \right) \psi' + \frac{e^A \psi''}{C} \right] = 0,
	\end{equation}
	
	\begin{equation}
		(h \sqrt{C} \phi')' =0.
	\end{equation}
	
	Integrating the last equation we obtain
	
	\begin{equation}
		\label{eqd}
		\phi' = \frac{\sqrt{\cal D}}{h \sqrt{C}}\ ,
	\end{equation}
	where $\cal D$ is a constant that represents the scalar charge of the phantom field and can be used to check the accuracy of numerical calculations. Its value as a function of frequency $\omega$ should be the same at different locations while fixing $r_0$ or $f_a$. By substituting the Eq.(\ref{eqd}) and the above Ansatze into the Einstein equations, with some combinations, we yields
	\begin{equation}
		- \frac{4 C^2 e^{-2A} \kappa \left( B \psi^2 \omega^2 + e^A \mu^2 \left( -1 + \sqrt{1 - 2B + 2B \cos\left[\frac{\psi}{f_a}\right]} \right) f_a^2 \right)}{B} + \frac{A'C'}{2} + C \left( \frac{2rA'}{h} + A'' \right) = 0 ,
	\end{equation}
	\begin{equation}
		- \frac{4 C^2 e^{-2A} \kappa \left( B \psi^2 \omega^2 + 2 e^A \mu^2 \left( -1 + \sqrt{1 - 2B + 2B \cos\left[\frac{\psi}{f_a}\right]} \right) f_a^2 \right)}{B} + \frac{3rC'}{h} - \frac{(C')^2}{2C} + C'' = 0 ,
	\end{equation}
	\begin{equation}
		\frac{2C h^2 \left[ \frac{r_0^2}{h^2} + \frac{C e^{-2A} \kappa \left( B \psi^2 \omega^2 + 2 e^A \mu^2 \left( -1 + \sqrt{1 - 2B + 2B \cos\left[\frac{\psi}{f_a}\right]} \right) f_a^2 \right)}{B} + \frac{(A')^2}{4} - \frac{rC'}{Ch} - \frac{(C')^2}{4C^2} + \kappa (\psi')^2 \right]}{\kappa} = {\cal D} .
	\end{equation}
	
	Together with Eq.(\ref{mat}) they form a system of second order ODEs to be solved numerically. Before going any further, we want to make two points: Firstly, when the axion field vanishes, one can derive the solution for an Ellis wormhole. Secondly, the throat parameter cannot be smoothly set to zero, preventing the model from reverting to the standard axion star solution. However, in the limit as this parameter approaches zero, the model's physical properties asymptotically converge to those of the standard axion star solution.
	
	In numerical calculations, we are particularly concerned about the results of some physical quantities, including the ADM mass of the gravitational system, can be read off directly from the asymptotic expansion of the metric component $g_{tt}$
	\begin{eqnarray}
		g_{tt}= -1+\frac{2 M}{r}+\cdots \ ,
	\end{eqnarray}
	and the Noether charge from the invariant under the $U(1)$ transformation of the axion field
	
	\begin{eqnarray}
		Q  &=& \int_{\cal S}J_s^t \nonumber \\
		&= &- \int J^t \left| g \right|^{1/2} dr d\Omega_{2},
	\end{eqnarray}
	with the conserved current
	\begin{equation}
		J^{\mu} = -i\left(\psi^*\partial^\mu\psi - \psi\partial^\mu\psi^*\right), \;\;\;\;\;\;\; J^\mu_{\,\,\,; \mu} =0 \;.
	\end{equation}
	
	In addition, we focus on the Kretschmann scalar $R^{\mu\nu\rho\sigma}R_{\mu\nu\rho\sigma}$ of this model, the expression is very long and not displayed here. The energy condition can be calculate from the combination of the components of the energy-momentum
	\begin{equation}
		\begin{split}
			\rho = -T^0_0 = - \frac{4 \mu^2 \left( -1 + \sqrt{1 - 2B + 2B \cos\left[\frac{\psi}{f_a}\right]} \right) f_a^2}{B} +\\ \frac{e^A \left( C^2 \left( -4r_0^2 - h^2A'^2 \right) + 4CrhC' + h^2C'^2 \right)}{4C^3\kappa h^2},
		\end{split}
	\end{equation}
	
	\begin{equation}
		\begin{split}
			\rho + P_1 = -T^0_0 + T^1_1= - \frac{4 \mu^2 \left( -1 + \sqrt{1 - 2B + 2B \cos\left[\frac{\psi}{f_a}\right]} \right) f_a^2}{B} +\\ \frac{e^A \left( C^2 \left( -4r_0^2 - h^2A'^2 \right) + 4CrhC' + h^2C'^2 \right)}{2C^3\kappa h^2}.
		\end{split}
	\end{equation}
	
	\section{BOUNDARY CONDITIONS}\label{sec3}

	Before numerically solving the differential equations instead of seeking the analytical
	solutions, we should provide appropriate boundary conditions. 
	
	Unlike in the study of boson stars with trivial topology, our analysis of wormhole spacetimes requires no restrictions at the origin. Instead, only need to satisfy the asymptotic flatness conditions
	
	\begin{eqnarray}
		\psi =A=0, \hspace{5pt}   C=1,\hspace{5pt}   
	\end{eqnarray}
	at infinity ($r \rightarrow \infty$).
	
	In this work, all the numbers are dimensionless as follows
	\begin{eqnarray}
		r \rightarrow r\mu \hspace{5pt}, \hspace{5pt} \psi \rightarrow \psi \kappa^{-1/2}\hspace{5pt}, \hspace{5pt} \omega \rightarrow \omega/\mu.
	\end{eqnarray}
	Without loss of generality, we can fix the specific parameters as $\mu = 1$ and $\kappa= 2$.
	To facilitate numerical calculations, we transform the radial coordinates by the following equation
	\begin{eqnarray}
		\label{transform}
		x= \frac{2}{\pi}\arctan(r) \;,
	\end{eqnarray}
	map the infinite region ($-\infty$,$+\infty$) to the finite region (-1,1).
	This allows the ordinary differential equations to be approximated by algebraic equations. The grid with 2000 points covers the integration region and the relative errors are less than $10^{-5}$.
	
	\section{NUMERICAL RESULTS}\label{sec4}
	
	\subsection{ADM mass and Noether charge}
	
	We select a representative set of parameters for our analysis: the decay constant is set to $f_a=\{1.0,0.12,0.10\}$, and the throat parameter to $r_0=\{0.0001,0.1,0.5,0.8\}$. The results are then grouped, with one parameter held constant in each case to investigate its specific influence. This range of parameters is sufficiently broad to explore the distinct physical properties of the model.
	
\begin{figure}
		\begin{center}
			\subfigure{\includegraphics[width=0.45\textwidth]{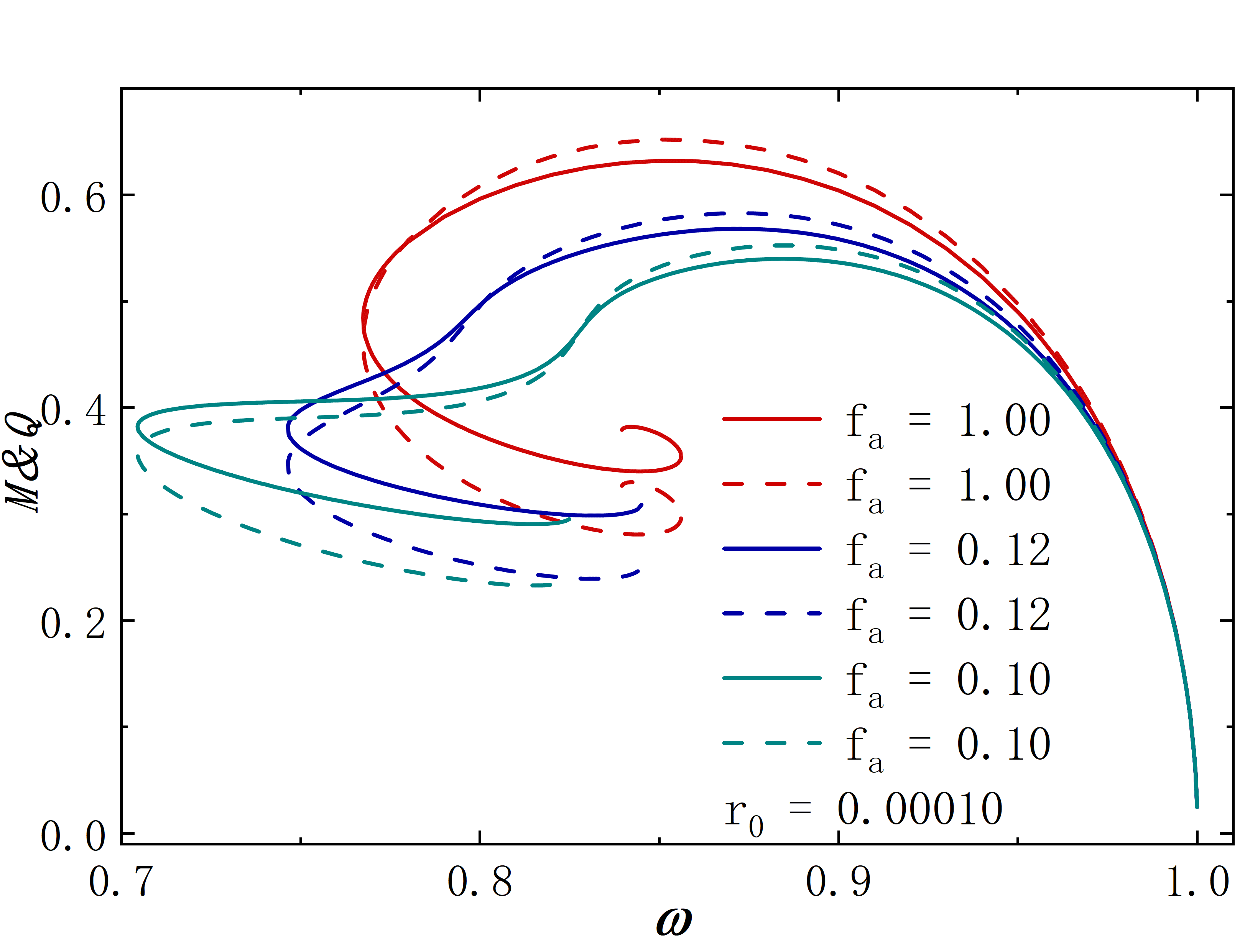}}
			\subfigure{\includegraphics[width=0.45\textwidth]{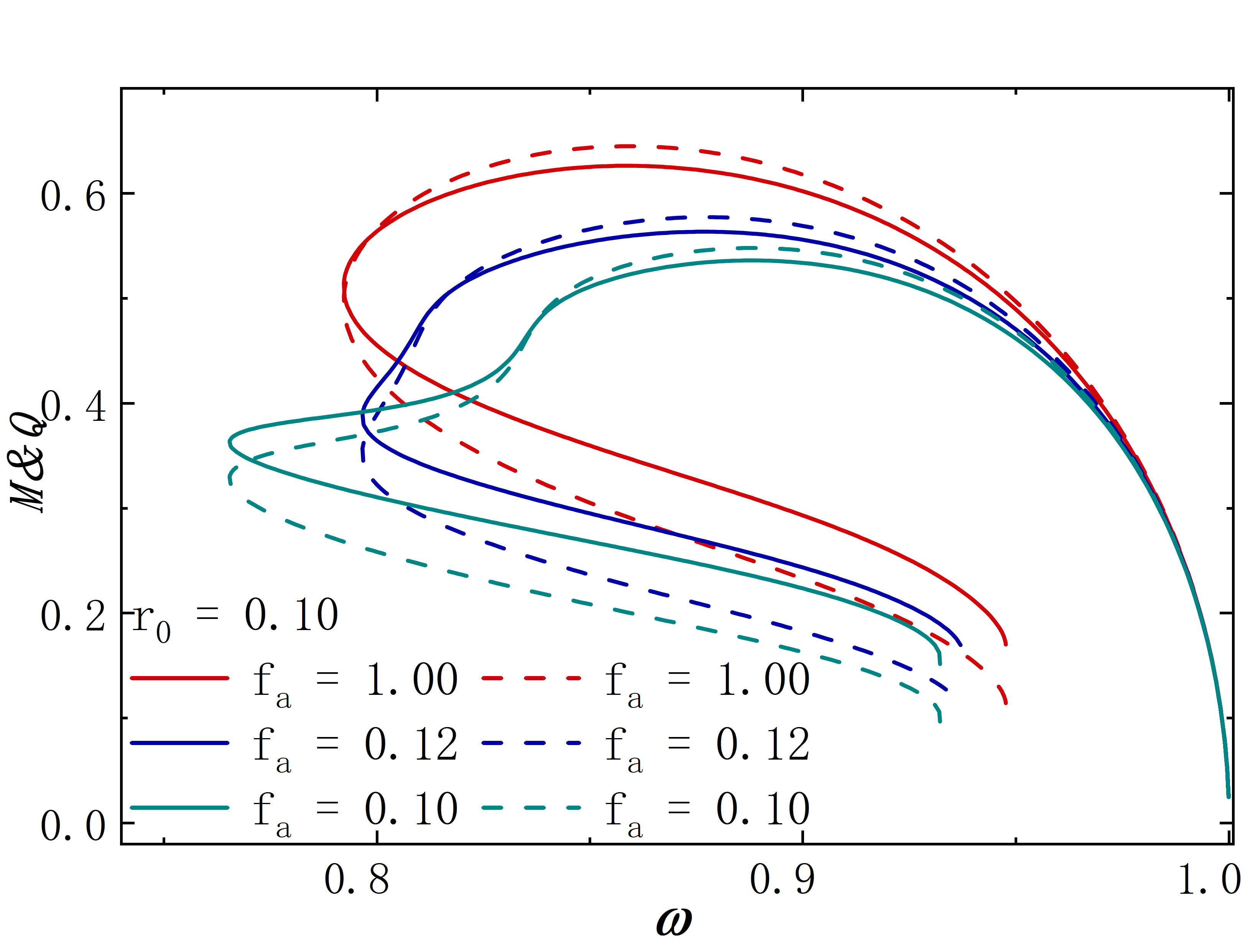}}
			\subfigure{\includegraphics[width=0.45\textwidth]{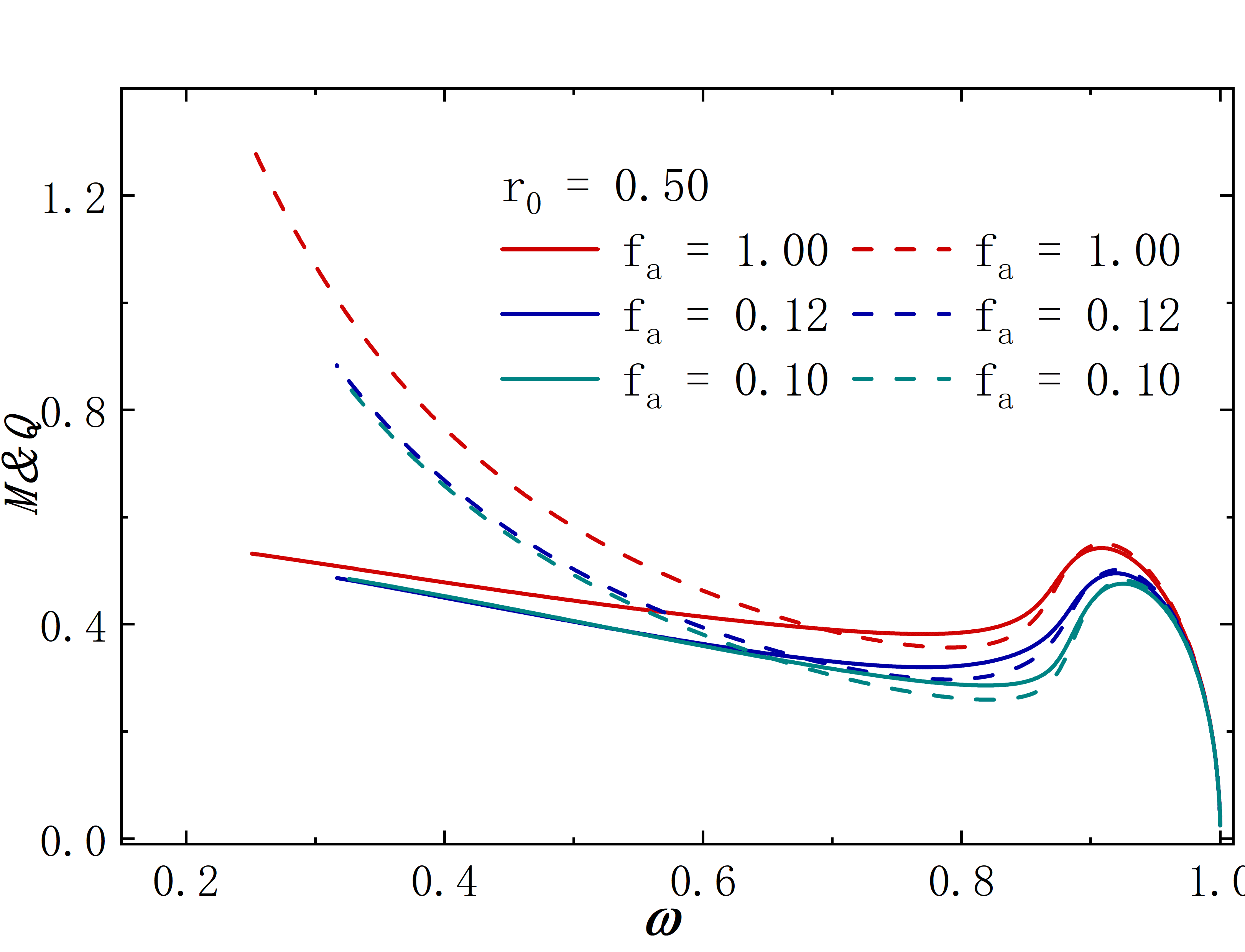}}
			\subfigure{\includegraphics[width=0.45\textwidth]{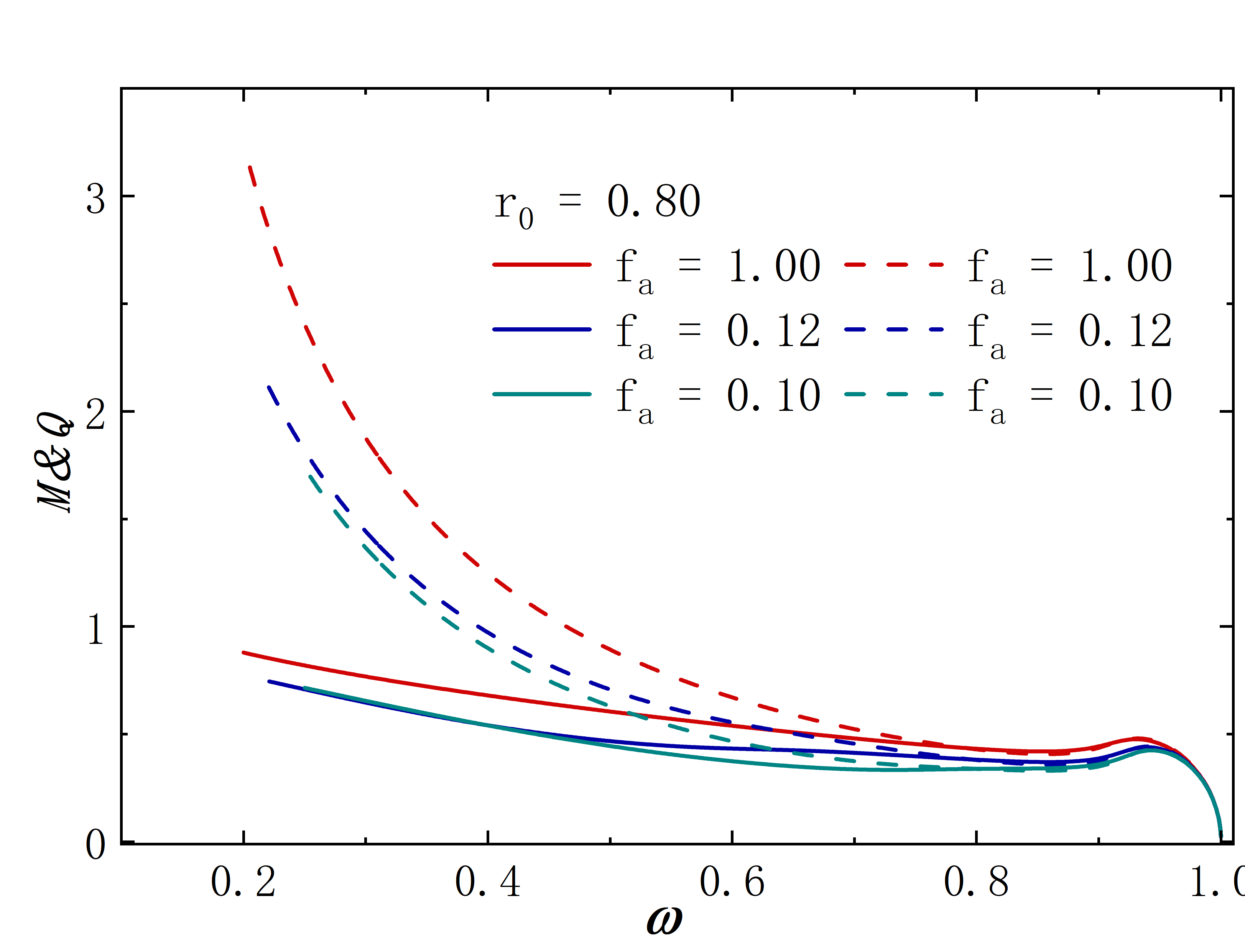}}
		\end{center}
		\caption{The ADM mass $M$ and Noether charge $Q$ as the function of frequency $\omega$ for some values of $f_a$ under $r_0 = 0.0001, 0.1, 0.5, 0.8$, the solid line represents $M$, and the dashed line represents $Q$.}
		\label{phase1}
	\end{figure}

	The domain of existence of the nontrivial topology axion star solutions for different decay constant $f_a$ in an ADM mass $M$ and Noether charge $Q$ vs. frequency $\omega$ diagram is shown in Fig.~\ref{phase1}. In the figure, the solid line represents $M$, and the dashed line represents $Q$.  When $f_a=1$, the curves of mass $M$ and charge $Q$ are very similar to those of boson stars with wormhole spacetime topology \cite{Dzhunushaliev:2014bya,Ding:2023syj}. This characteristic persists under different values of the throat parameter. Specifically, the curves exhibit a spiral shape with a tight winding at smaller values of the $r_0$, which opens up as the $r_0$ increases. This behavior suggests that in the large values of $f_a$, the axion star solution degenerates into a mini-boson star model. It is noteworthy that as the parameter $f_a$ decreases, when $r_0$ is smaller, the curve exhibits a ``duckbill" shape. In contrast, when $r_0$ is large, the curve maintains a spiral expansion.
	
	From the perspective of catastrophe theory and the numerically evolved analysis \cite{Lee:1988av,Gleiser:1988rq,Herdeiro:2021lwl} axion stars with smaller $f_a$, exhibit two stable branches separated by an unstable region. However, it remains to be seen whether these conclusions hold in the context of wormhole spacetimes. To facilitate a future stability analysis and data extraction, we present in \ref{tab1} the point corresponding to the maximum mass on the first branch for various $r_0$ solutions under two distinct sets of the parameter $f_a$. In a trivial spacetime, this specific point delineates the boundary between stable Newtonian and unstable branches. In addition to the ADM mass and the corresponding frequency, we also show the numerical value of the axion scalar field $\psi_0$ at $x=0$ at this time.

	\begin{table}[htbp]
		\centering
		\renewcommand{\arraystretch}{1.4}
		\setlength{\tabcolsep}{14pt}
		\begin{tabular}{|c|c|c|c|c|}
			\hline
			\multicolumn{2}{|c|}{} & $\psi_0$ & $\omega$ & $M$ \\ \hline
			\multirow{3}{*}{$f_a=1.0$} & $r_0=0.0001$ &   0.19110       &  0.85350        &  0.63209        \\ \cline{2-5}
			& $r_0=0.1$   &  0.18588        &   0.85920       &   0.62623       \\ \cline{2-5}
			& $r_0=0.5$   &  0.12895        &  0.90790        &  0.54220        \\ \hline
			\multirow{3}{*}{$f_a=0.12$} & $r_0=0.0001$ &  0.17391        & 0.87260         &  0.56801        \\ \cline{2-5}
			& $r_0=0.1$   & 0.16979         &  0.87700        &   0.56340       \\ \cline{2-5}
			& $r_0=0.5$   &  0.11580        &  0.91950        &  0.49510        \\ \hline
		\end{tabular}
		\caption{The parameters $\psi_0$, $\omega$, $M$ at the point corresponding to the maximum mass on the first branch for various $r_0$ solutions under $f_a = 1.0, 0.12$.}
		\label{tab1}
	\end{table}
	
	\subsection{Phantom scalar charge and The Kretschman scalar}
	
	The nature of a wormhole is intimately linked to its throat parameter $r_0$. In particular, the magnitude of $r_0$ directly influences the wormhole's traversability. While a complete verification of traversability typically necessitates the full numerical evolution of a test particle passing through the wormhole \cite{Kain:2023ore}, for static spherically symmetric solutions, analyzing the Kretschmann scalar at the throat suffices to address this issue.
	
\begin{figure}
		\begin{center}
			\subfigure{\includegraphics[width=0.45\textwidth]{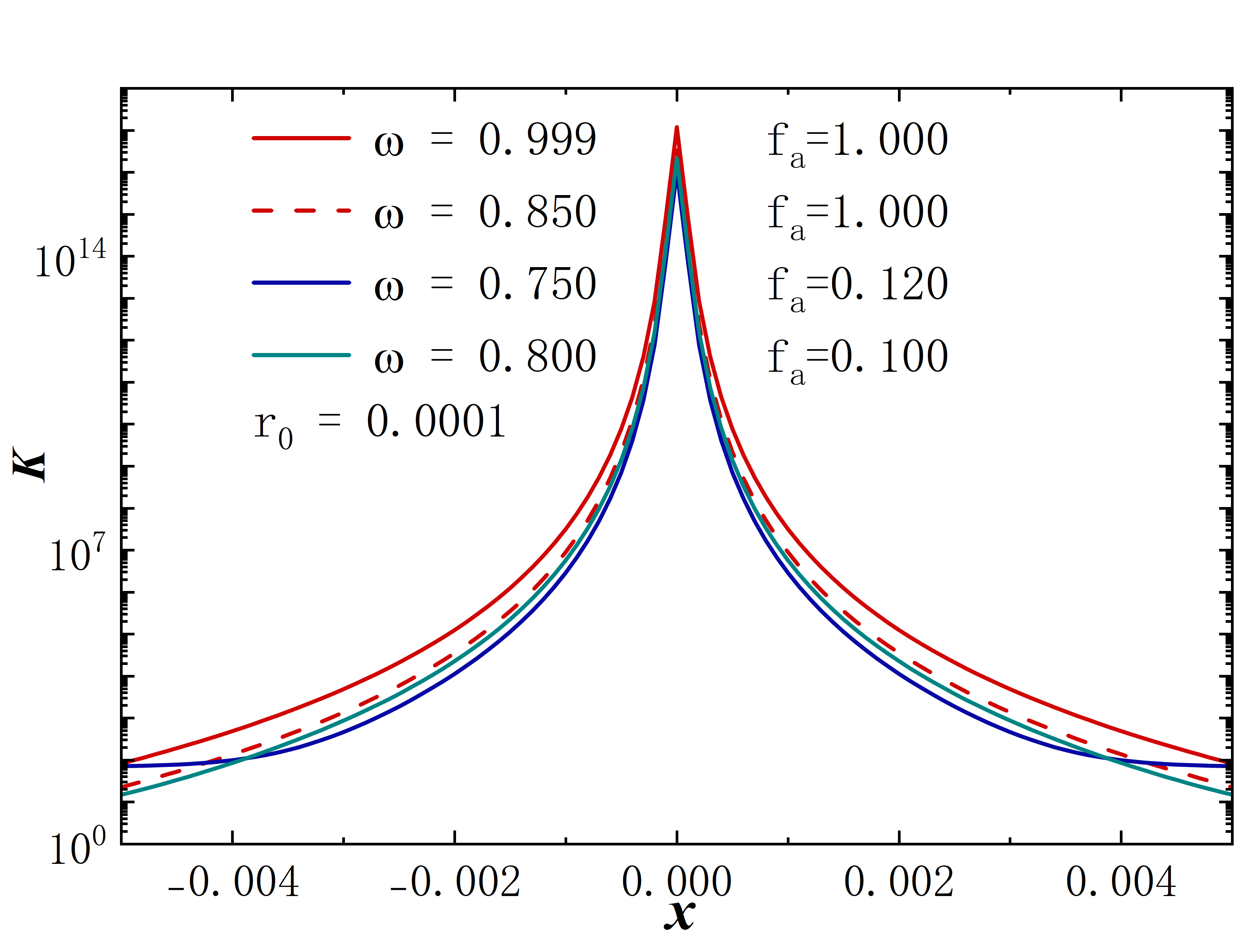}}
			\subfigure{\includegraphics[width=0.45\textwidth]{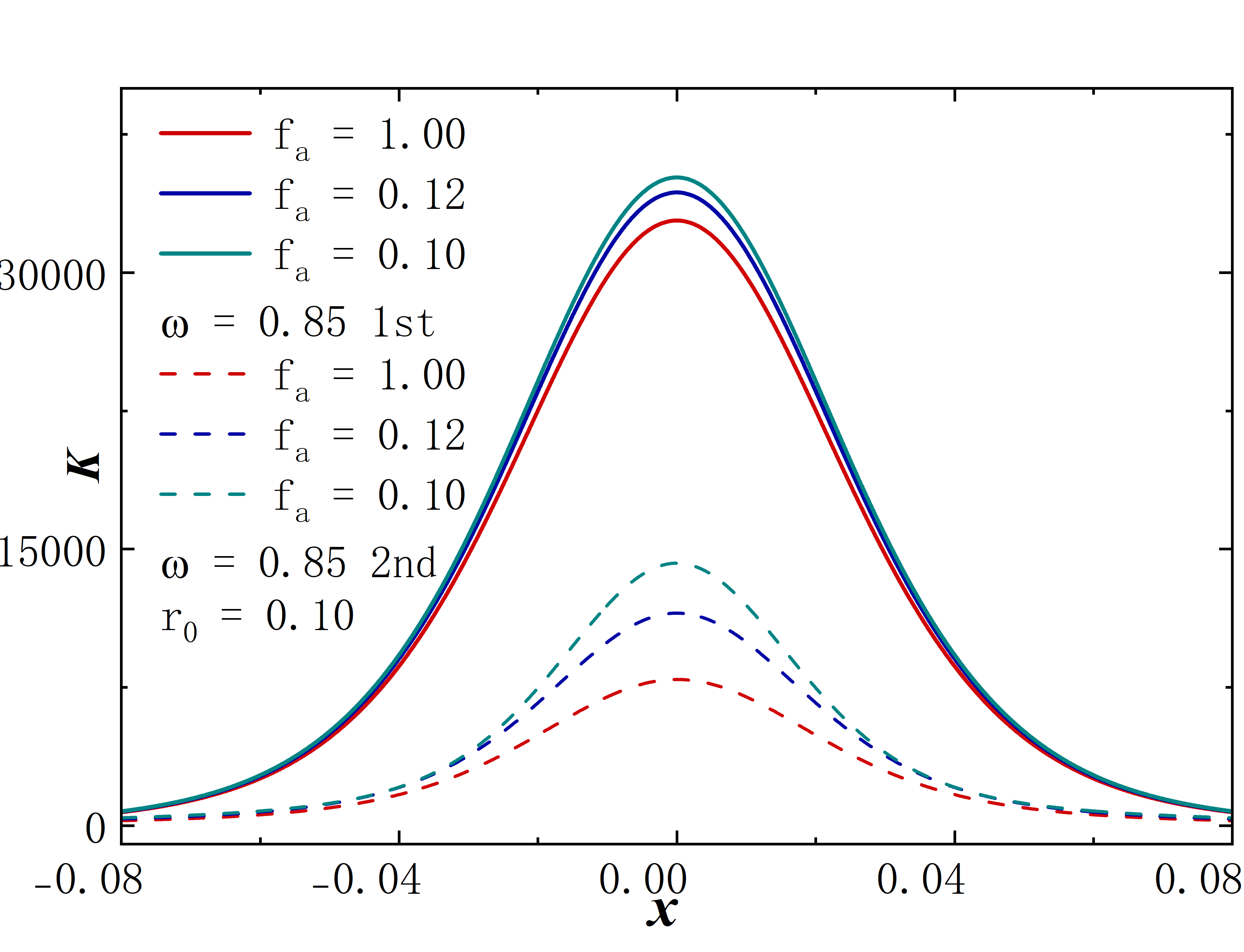}}
			\subfigure{\includegraphics[width=0.45\textwidth]{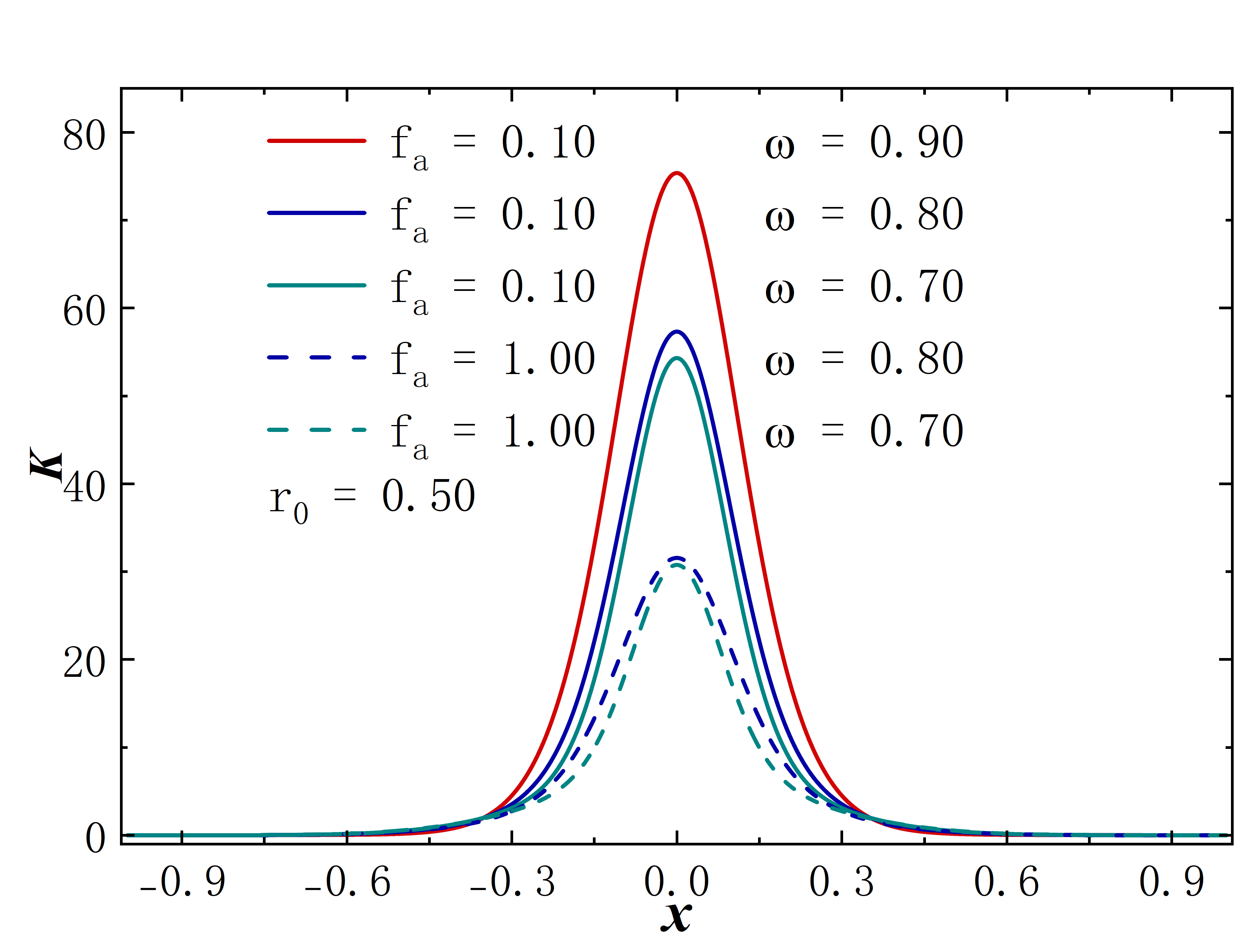}}
			\subfigure{\includegraphics[width=0.45\textwidth]{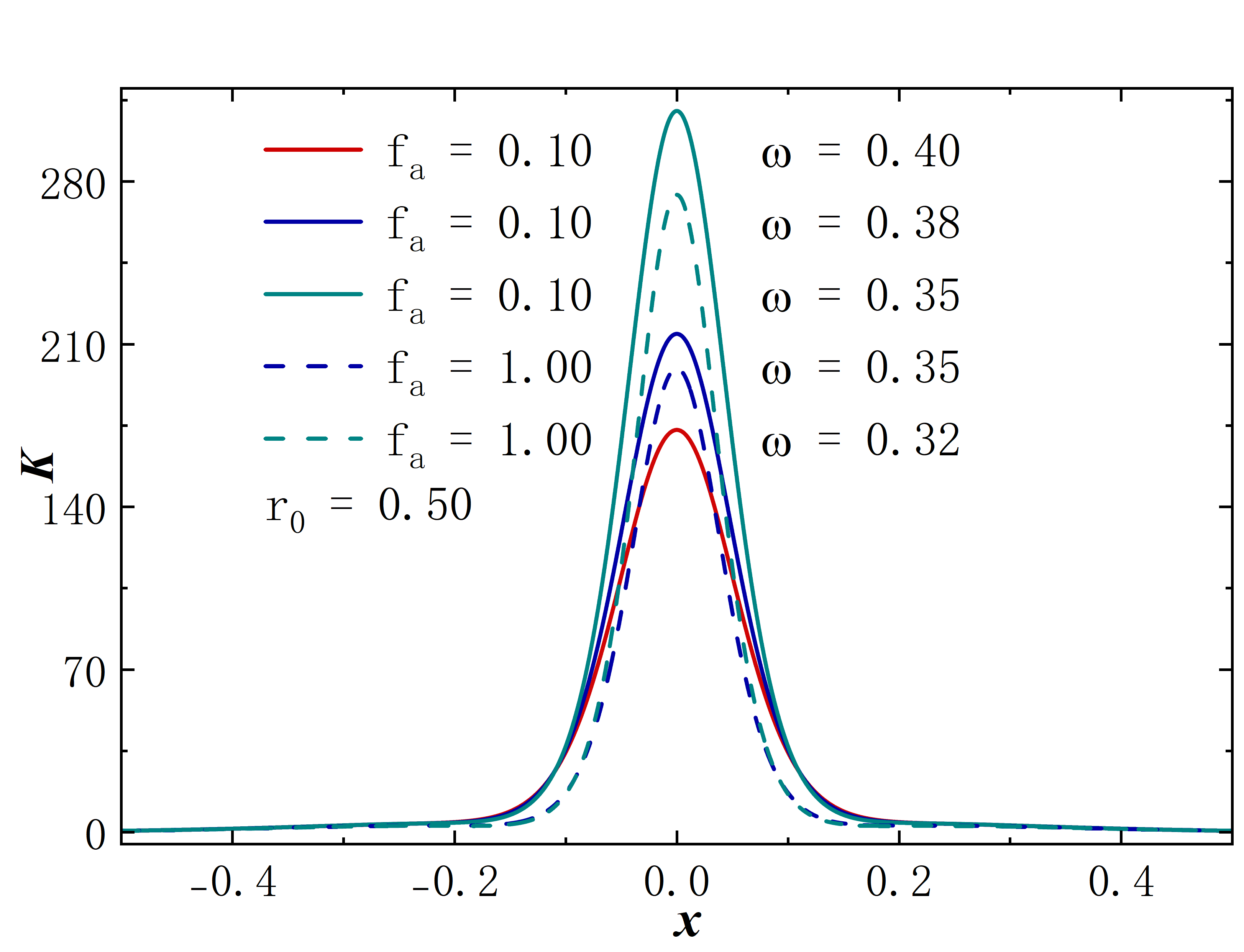}}
		\end{center}
		\caption{The Kretschmann scalar vs. radial coordinate $x$ under different values of $r_0$ and $f_a$.}
		\label{phase2}
	\end{figure}

	For a throat parameter of $r_0=0.0001$, the Kretschmann scalar $K$ is exceptionally large, exhibiting a divergent trend near the throat, irrespective of the values of the $f_a$ and frequency $\omega$. A consistent observation is that as $r_0$ increases, the overall value of the Kretschmann scalar progressively decreases.  In addition to this direct dependence on $r_0$, $K$ is also systematically influenced by $f_a$ and $\omega$. We show the diagram of $K$ as a function of the radial coordinate $x$ in Fig.~\ref{phase2}.
	
\begin{figure}
		\begin{center}
			\subfigure{\includegraphics[width=0.45\textwidth]{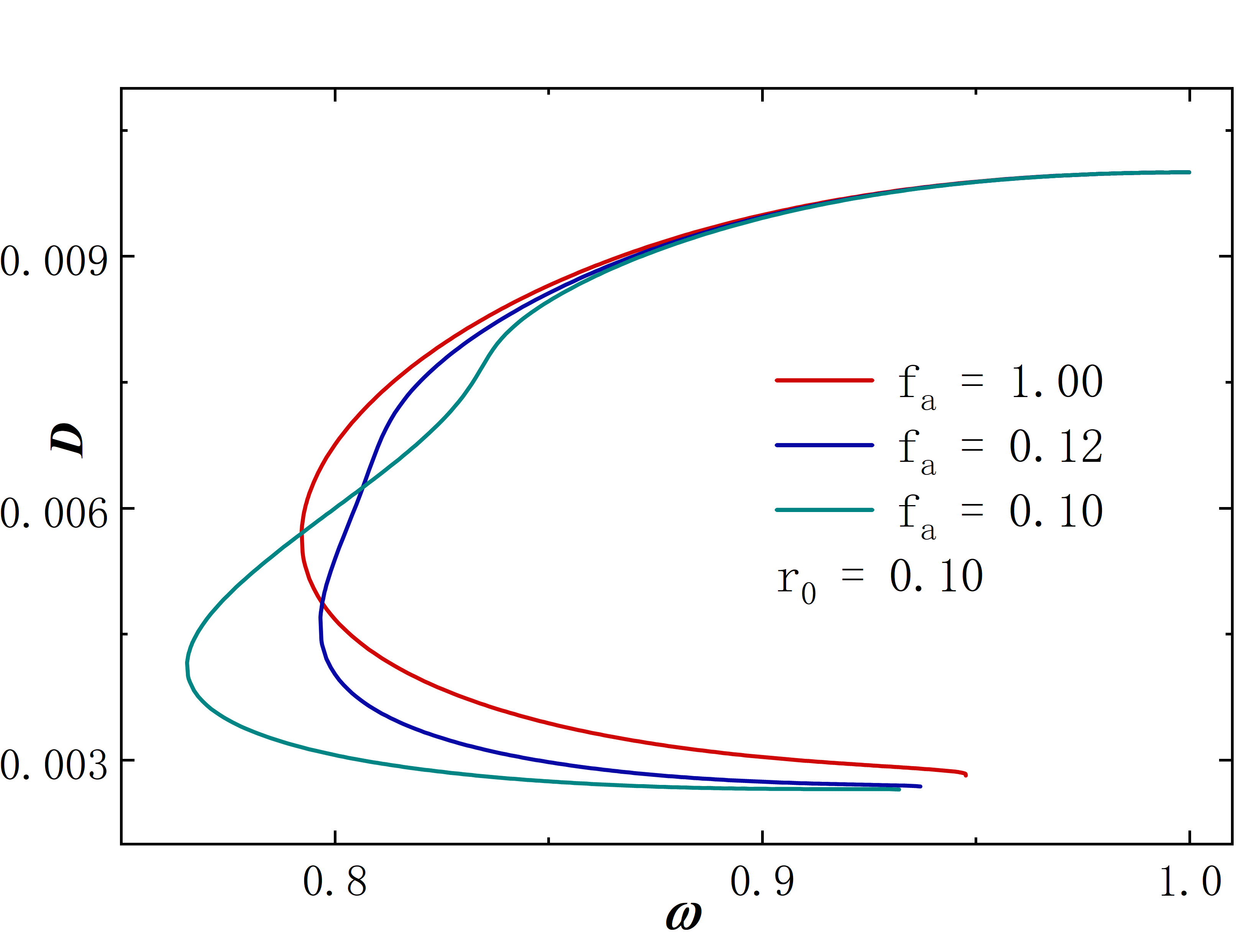}}
			\subfigure{\includegraphics[width=0.45\textwidth]{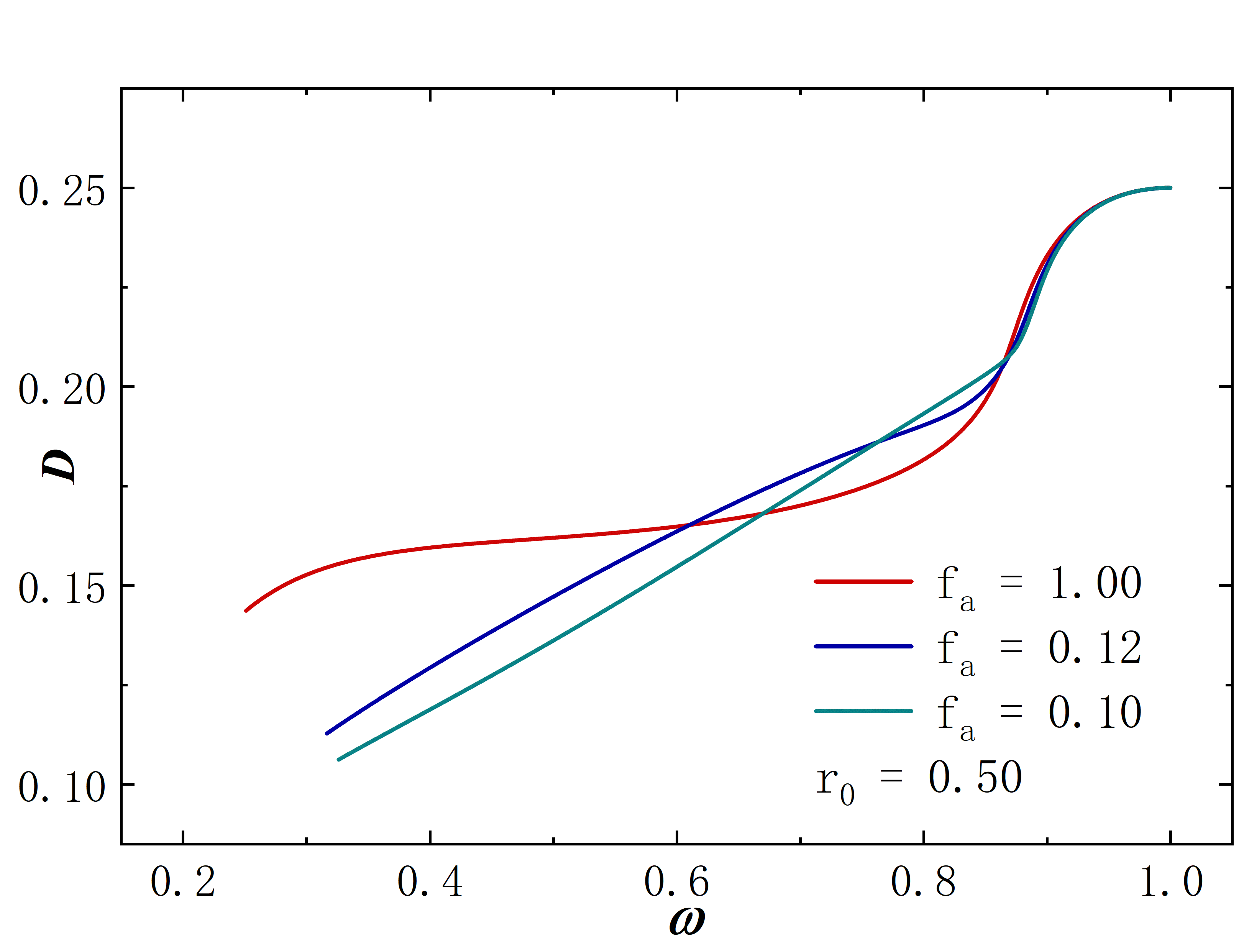}}
			\subfigure{\includegraphics[width=0.55\textwidth]{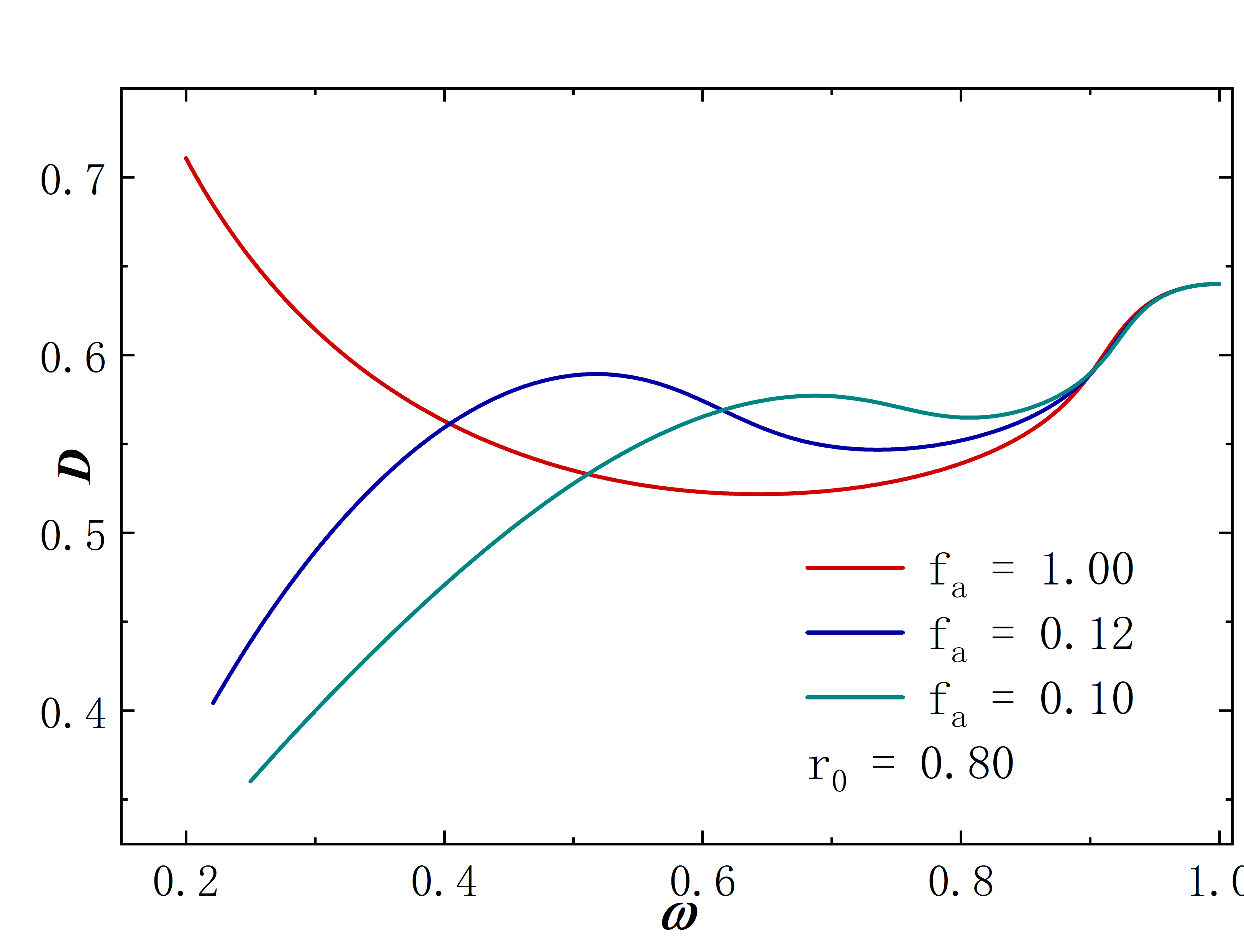}}
		\end{center}
    		\caption{The phantom field scalar charge $\cal D$ vs. field frequency $\omega$ under different values of $r_0$ and $f_a$.}
		\label{phase3}
	\end{figure}

            	Fig.~\ref{phase3} shows the distribution of a physical quantity closely related to the wormhole throat parameter: the phantom field scalar charge $\cal D$. The value of this quantity serves as a measure for the phantom field content within the model. We find that the scalar charge exhibits a positive correlation with the throat parameter $r_0$, while the parameter $f_a$ has relatively little effect on it.
	
	In summary, for extremely small throat parameters $r_0$, the value of the parameter $\cal D$ indicates a scarcity of the phantom field, while the corresponding Kretschmann scalar $K$ is of a high order of magnitude, signifying a violation of traversability. As $r_0$ increases, $\cal D$ also increases while $K$ generally decreases. However, in this regime, the value of $K$ is also influenced by the parameters $f_a$ and frequency.

	\subsection{Energy Condition}
	
	Generally, a wormhole model supported by a phantom field violates the null energy condition, and consequently the weak and strong energy conditions. However, this situation may change in certain parameter regimes when other matter fields are also coupled to the system.
	
\begin{figure}[H]
		\begin{center}
			\subfigure{\includegraphics[width=0.45\textwidth]{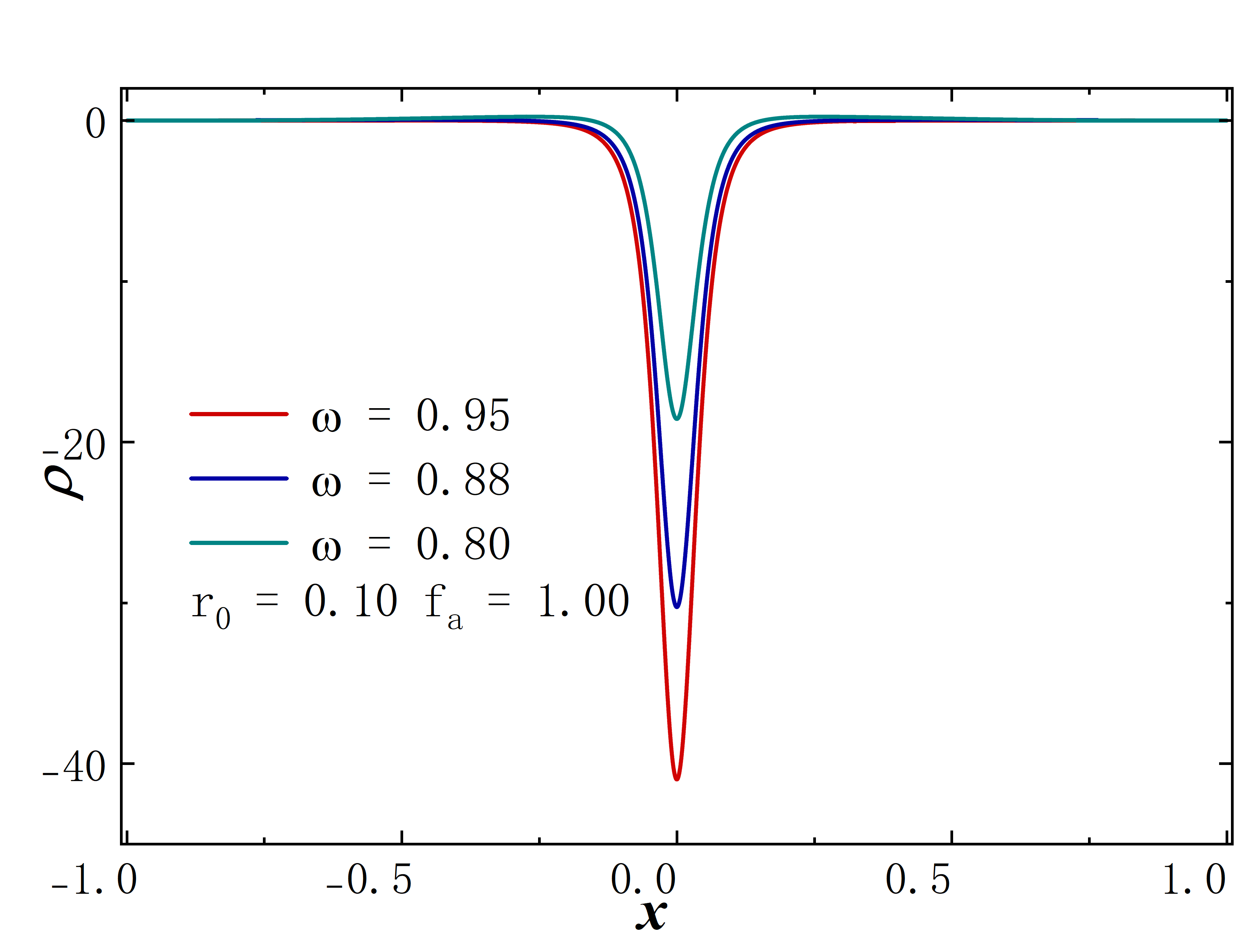}}
			\subfigure{\includegraphics[width=0.45\textwidth]{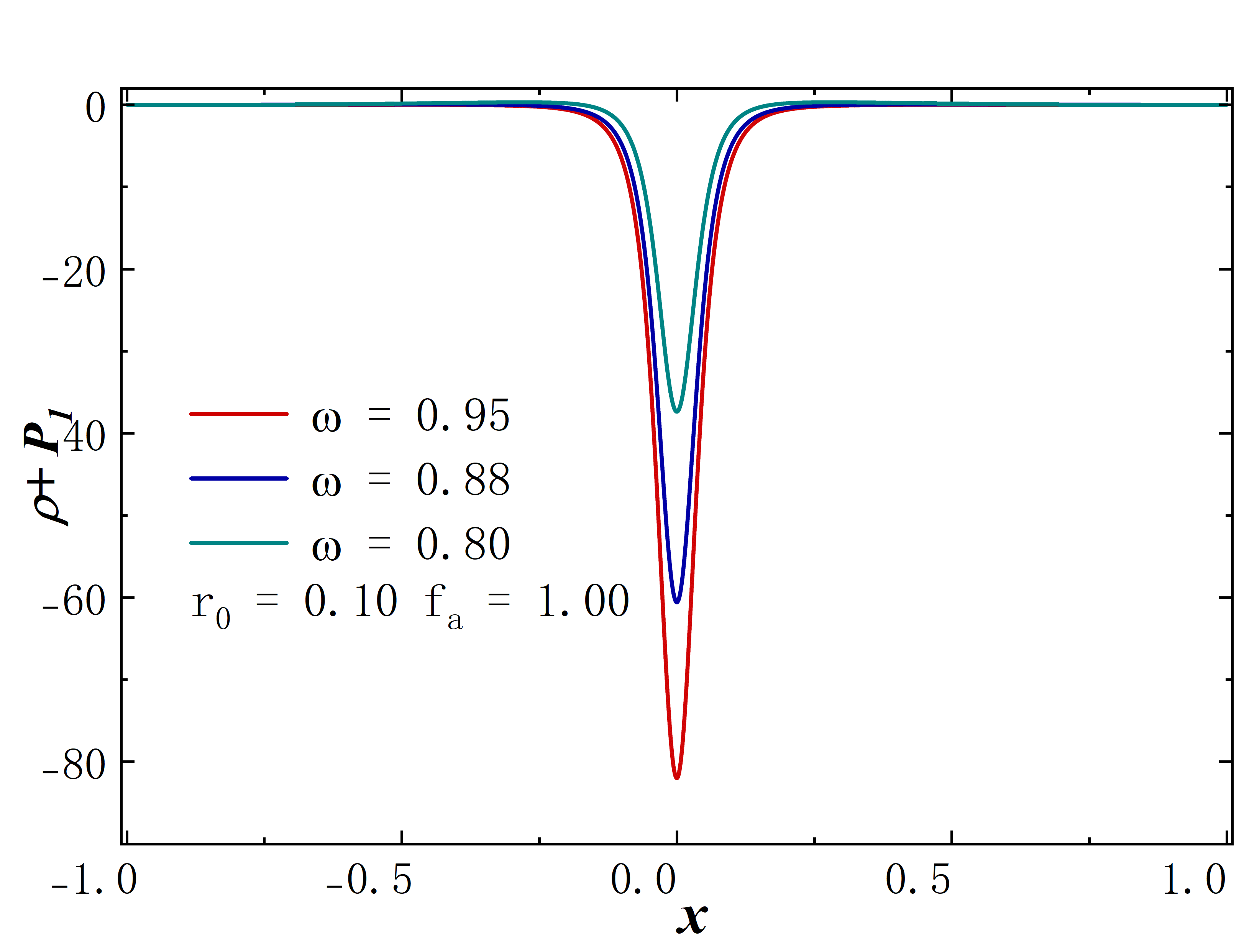}}
			\subfigure{\includegraphics[width=0.45\textwidth]{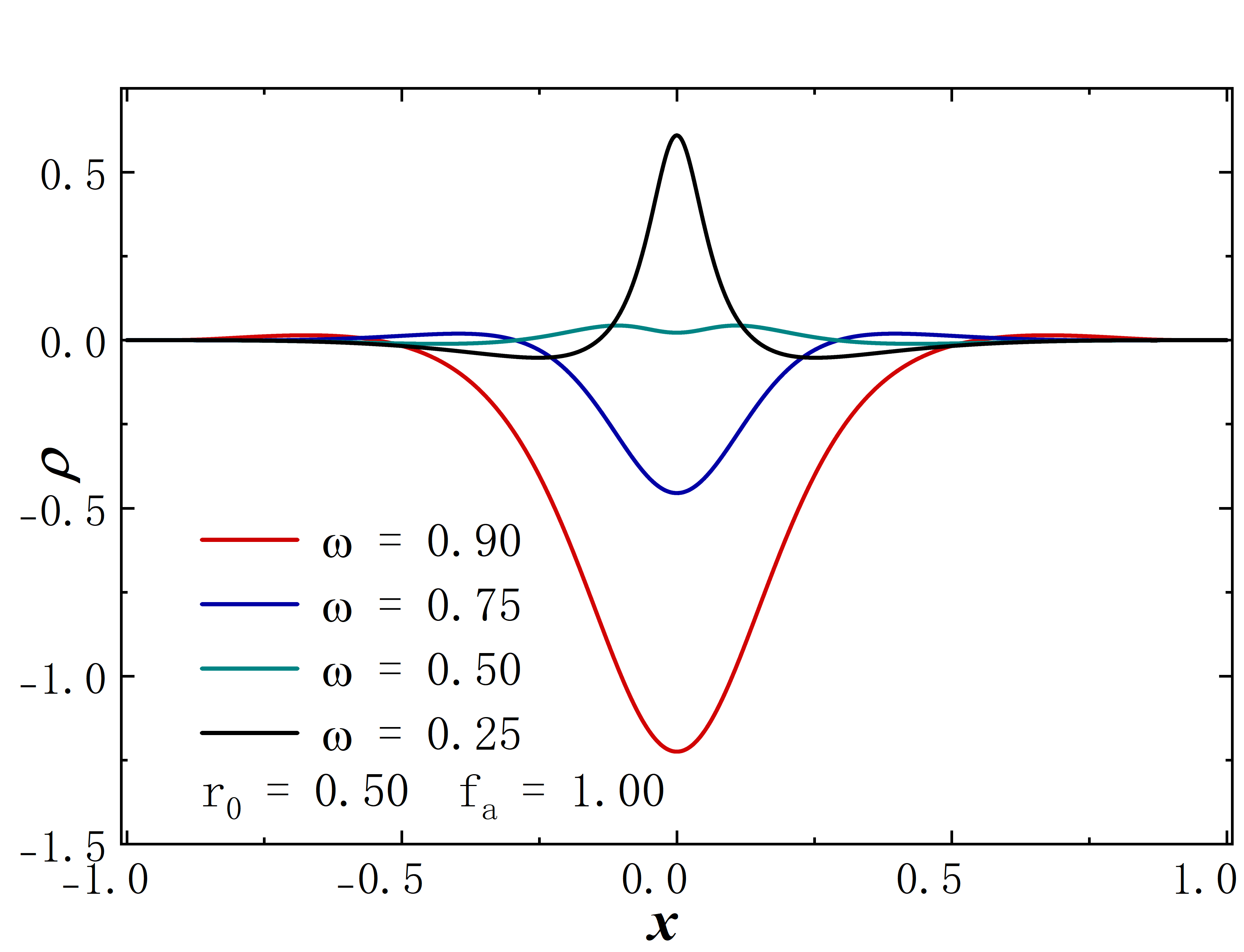}}
			\subfigure{\includegraphics[width=0.45\textwidth]{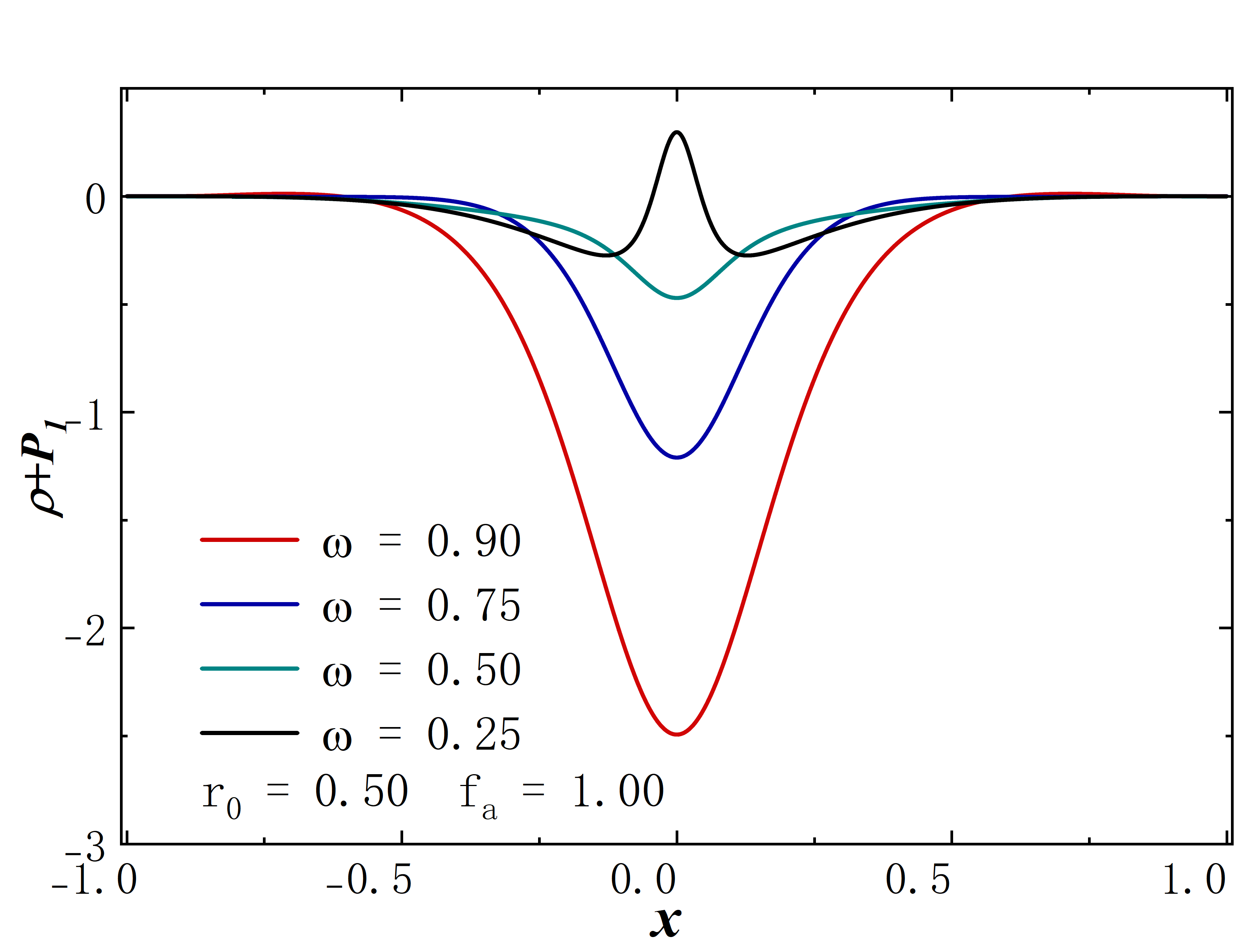}}
			\subfigure{\includegraphics[width=0.45\textwidth]{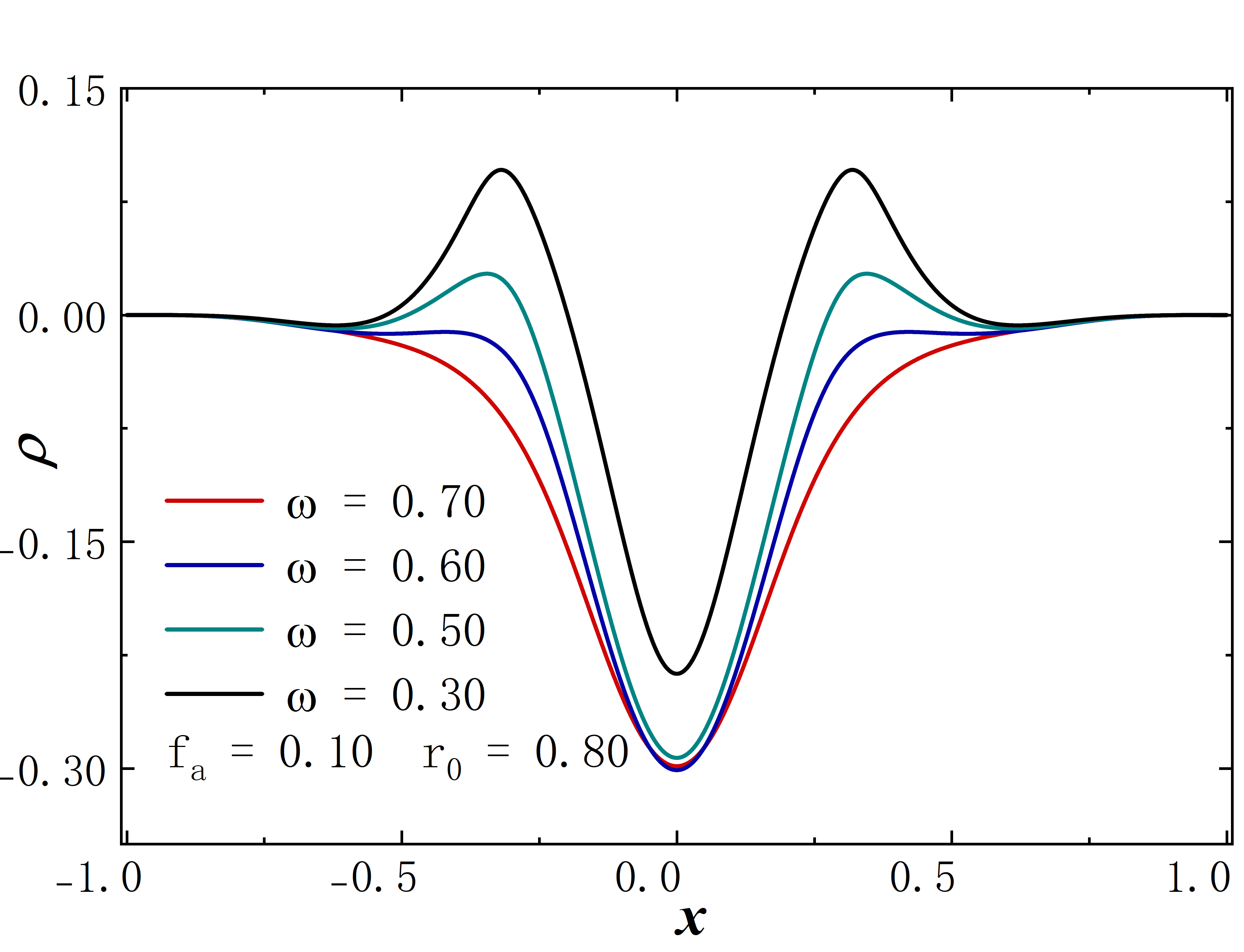}}
			\subfigure{\includegraphics[width=0.45\textwidth]{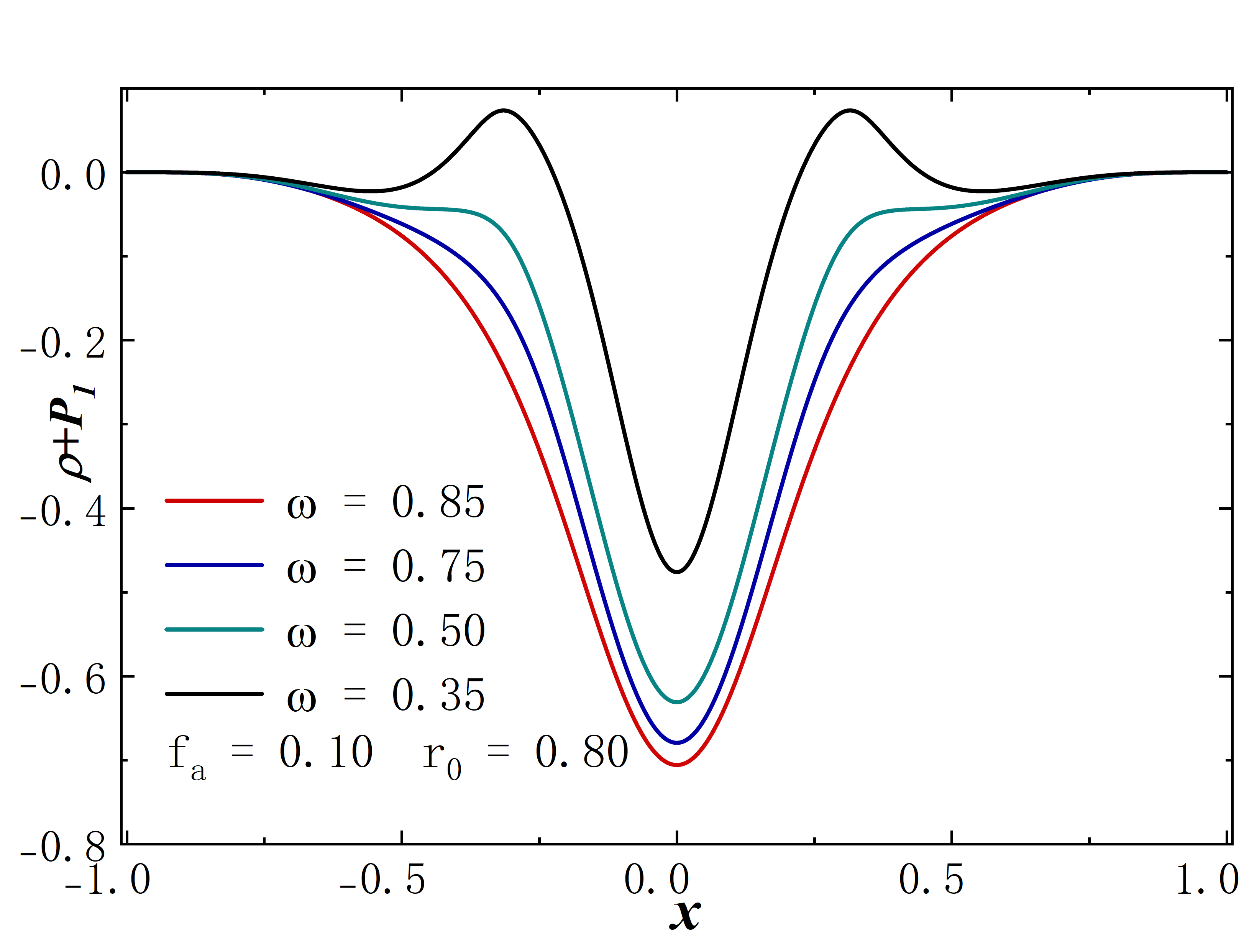}}
		\end{center}
		\caption{Energy density $\rho$ and the sum of energy density and radial pressure $\rho+p_r$ for various throat parameters $r_0$ and frequency $\omega$.}
		\label{phase4}
	\end{figure}

	In Fig.~\ref{phase4}, we show the distribution of the system's energy density $\rho$ and the sum of energy density and radial pressure $\rho+p_r$ for various throat parameters $r_0$ and frequency $\omega$. Since the parameter $f_a$ has a negligible impact on the results, we have selected only one representative value of $f_a$ for each group.
	
	A more interesting situation arises for relatively large throat parameters, specifically for the cases of $r_0=0.5$ and $r_0=0.8$. In the case of $r_0=0.5$, we observe that as the frequency decreases, both the $\rho$ and the $\rho+p_r$ progressively increase. For a sufficiently small frequency $\omega$, these quantities become positive, indicating that the null energy condition (NEC) is no longer violated throughout the entire spacetime. In contrast, for $r_0=0.8$, while the general trends described above persist, the NEC at the throat remains violated even in the low-frequency limit. Interestingly, two symmetric regions emerge near the throat where the NEC is satisfied.
	This phenomenon is, of course, a result of the combined contribution of the axion scalar field and the phantom field. The former provides positive energy density, while the latter supplies the necessary negative energy. However, this behavior is not universal across all models. The properties of gravitational configurations formed by mixing various normal matter fields with phantom fields are governed by the interplay between these two types of ``normal" and ``exotic" matter, leading to a rich variety of complex possibilities.
	
	\subsection{Wormhole Geometries}
	
	We now turn to the geometry of the wormholes within the axion boson stars. We can make use of a geometrical embedding diagram by fixing $t$ and $\theta$. The resulting two-dimensional spatial hypersurface of the wormhole spacetime can then be embedded in a three-dimensional Euclidean space, where the embedding diagram can be used to visualize the wormhole geometry. This technique allows us to better understand the topology and properties of the wormhole solution.

	The specific method is that we begin by constructing the embeddings of planes with  $\theta = \pi/2$, and then use the cylindrical coordinates $(\rho,\varphi,z)$, the metric on this plane can be expressed by the following formula
	\begin{align}
		\nonumber ds^2 &= C e^{-A}  d r^2 + C e^{-A} h   d\varphi^2 \, \\
		&= d \rho^2 + dz^2 + \rho^2 d \varphi^2   \,.
	\end{align}
	Comparing the two equations above, we then obtain the expression for $\rho$ and $z$,
	\begin{equation} \label{formula_embedding}
		\rho(r)= \sqrt{ C(r) e^{-A(r)} h(r) } ,\;\;\;\;\;\;\;\;\;\;   z(r) = \pm  \int  \sqrt{ C(r) e^{-A(r)}  -   \left( \frac{d \rho}{d r} \right)^2    }     d r \;.
	\end{equation}
	Here $\rho$ corresponds to the circumferential radius, which corresponds to the radius of a circle located in the equatorial plane and having a constant coordinate $r$. The function $\rho(r)$ has extreme points, where the first derivative is zero.
	When the second derivative of the extreme point is greater than zero, we call this point a throat, which corresponds to a minimal surface. When the second derivative of the extreme point is less than zero, we call this point an equator, which corresponds to a maximal surface.
	
	Due to the solution's symmetry with respect to the origin, the wormhole exhibits one of two possible topological configurations: a single throat centered at $z=0$, or a pair of throats situated symmetrically on either side of a equatorial plane which is located at $z=0$. We found that for relatively small values of the throat parameter $r_0$, the wormhole consistently has a single throat and no equatorial plane, irrespective of the values of the axion decay constant $f_a$ and frequency $\omega$.
	
\begin{figure}
		\begin{center}
			\subfigure{\includegraphics[width=0.4\textwidth]{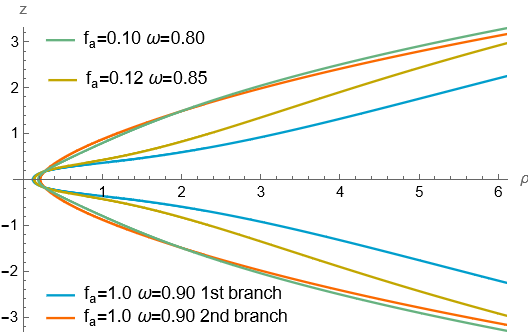}}
			\subfigure{\includegraphics[width=0.4\textwidth]{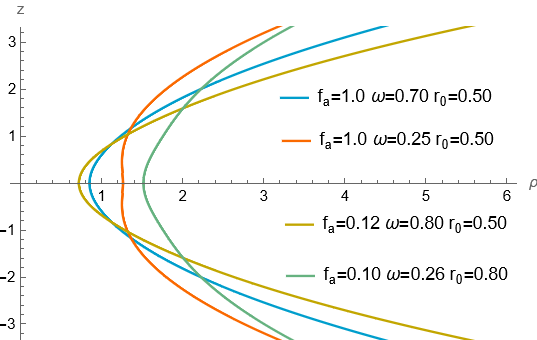}}
			\subfigure{\includegraphics[width=0.8\textwidth]{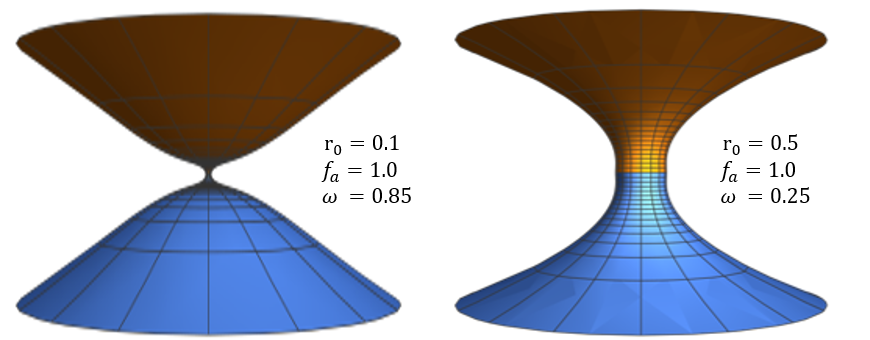}}
		\end{center}
		\caption{Two-dimensional view of the isometric embedding of the equatorial plane and the corresponding 3D embedding diagrams of wormhole solutions for different $f_a$ and $\omega$ with $r_0 = 0.1 $ in the left, and $r_0 = 0.5, 0.8$ with different $f_a$ and $\omega$ on the right. } 
		\label{phase5}
	\end{figure}

	A more complex behavior emerges as $r_0$ increases to a value that allows for very small frequencies. For instance, when $f_a=1$, the wormhole has a single throat at large frequencies. However, as $\omega$ becomes very small, a single throat gives way to an equatorial plane appearing at the wormhole's center. In contrast, for small values of $f_a$, such as $0.12$ or $0.1$, the single-throat structure is maintained at all frequencies. We present the corresponding two sets of wormhole embedding diagrams in Fig.~\ref{phase5}.
	
	\subsection{Null geodesics and Light Ring}
	
	Finally, we analyze the null geodesics and the resulting light ring within the model's equatorial plane, a simple yet physically significant configuration. Refs. \cite{Cunha:2022gde,Xavier:2024iwr} have separately calculated the light ring characteristics for extremely compact relativistic stars such as boson stars, and for wormholes. Given that our model can be viewed as a wormhole spacetime with axion star properties, its light ring structure should resemble that of a general axisymmetric stationary wormhole.
	
	Photons in a gravitational field move along null geodesics 
	\begin{equation} 
		g_{\mu \nu} \dot{x}^{\mu} \dot{x}^{\nu}=0,
	\end{equation}
	and in our case $x^{\mu}=\left ( t,r,\theta ,\varphi  \right ) $, the dots represent derivatives with respect to the affine parameter $\lambda$ along the geodesics. Due to the static character and the spherical symmetry of the system, we can assume that the orbit lies in the equatorial plane $\theta = \pi/2$ and defines two conserved quantities: Conserved energy $E=-g_{tt} \dot{t}$ and the angular momentum $L=g_{\varphi\varphi} \dot{\varphi}$. Taking into account the relevant metric components, the geodesic equation can be written as (re-scaling the affine parameter as $\lambda=\lambda/L$)
	
	\begin{equation}
		\dot{r}^{2}+\frac{1}{g_{tt}} \frac{1}{g_{rr}}(\frac{1}{b^{2}}+\frac{g_{tt}}{g_{\varphi \varphi }})=\dot{r}^{2}  +\frac{1}{C} (\frac{e^{A}}{Ce^{-A}h}-\frac{1}{b^{2}})=0,
	\end{equation}
	where the impact parameter as $b\equiv \frac{L}{E}  $  for a certain null particle. We define the effective potential $V_{eff}$ as
	
	\begin{equation}
		V_{eff}=\frac{e^{A}}{Ce^{-A}h},
	\end{equation}
	where the extreme points $r_{LR}$ satisfy 
	\begin{equation}
		\left.\frac{d V_{e f f}}{d r}\right|_{r_{L R}}=0.
	\end{equation}
	
	\begin{figure}
		\begin{center}
			\subfigure{\includegraphics[width=0.45\textwidth]{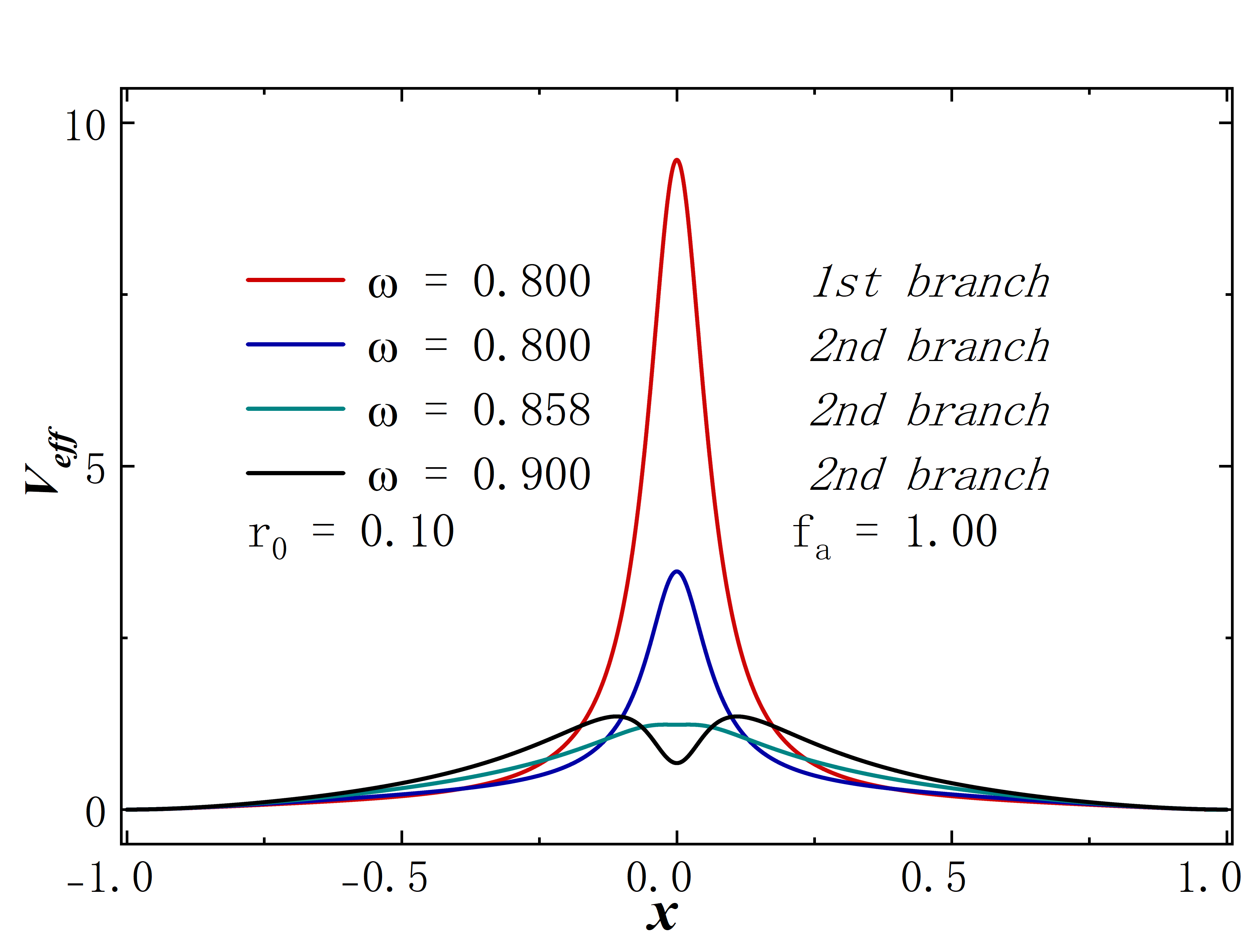}}
			\subfigure{\includegraphics[width=0.45\textwidth]{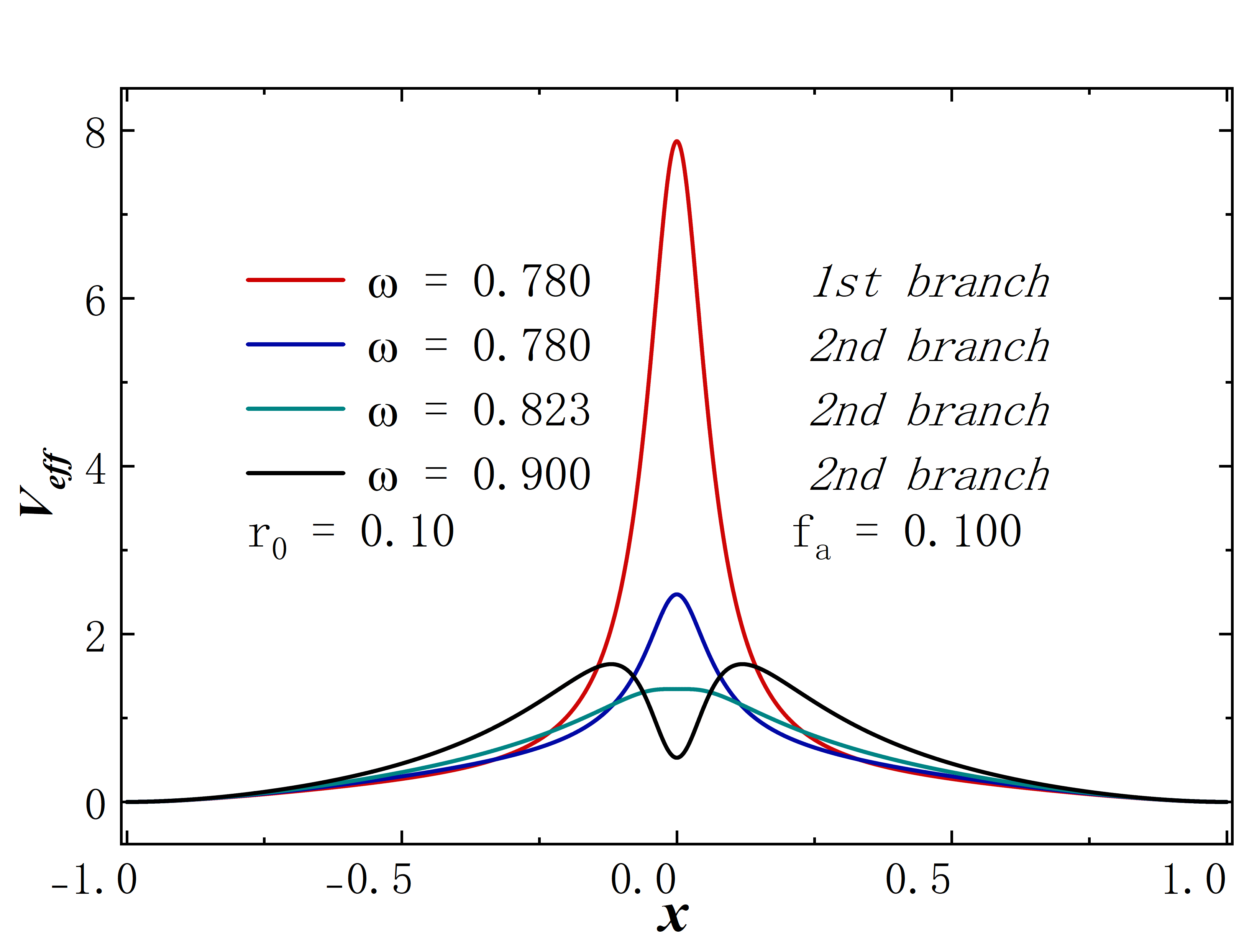}}
			\subfigure{\includegraphics[width=0.45\textwidth]{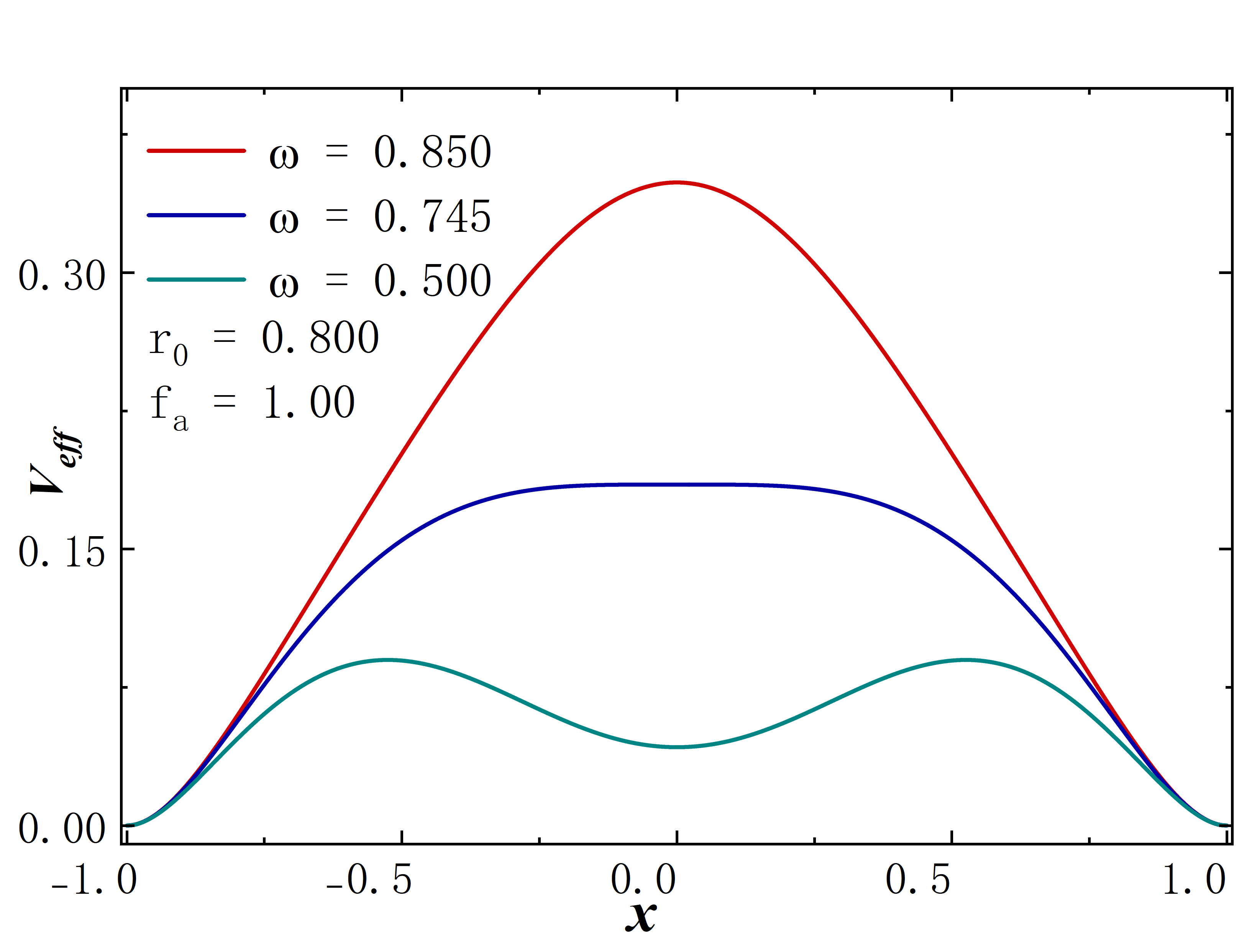}}
			\subfigure{\includegraphics[width=0.45\textwidth]{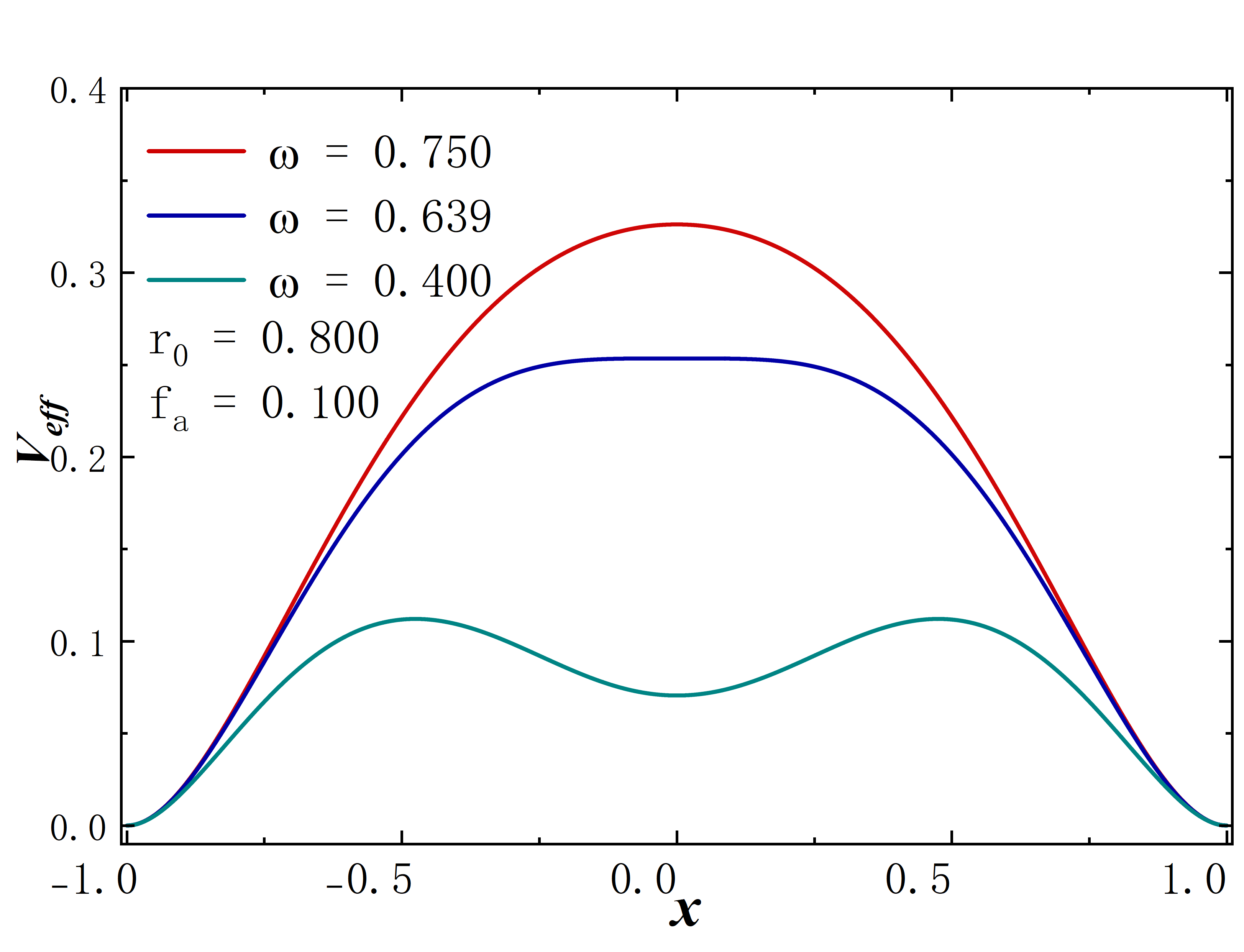}}
		\end{center}
		\caption{Effective potential as a function of radius $x$ with different parameters $\omega$, $r_0$ and $f_a$.}
		\label{phase6}
	\end{figure}
	
	For a null particle, if the square of the impact parameter $b_{c}^{2}=C(r_{L R})e^{-A(r_{L R})}h / e^{A(r_{L R})},$ photons coming from elsewhere will reduce the radial velocity to 0 at $r_{L R}$ and rotate around the wormhole. The corresponding orbits of the null particle are then referred to as the light ring. Moreover, for the maximum point of the effective potential, $V_{eff}^{''}(r_{LR})<0$, the light ring corresponding to $r_{LR}$ is unstable. On the contrary, the minimum point of the effective potential corresponds to a stable light ring.

	In the Fig.~\ref{phase6}, we find that the effective potential exhibits an extreme value at $x=0$, which may correspond to either a stable or an unstable orbit, regardless of the values of the parameters. In certain parameter regimes, multiple extreme points can emerge throughout the wormhole spacetime. For instance, as shown in the figure, three such extreme points are observed. The number of unstable orbits is always one greater than the number of stable orbits, a finding that is consistent with the conclusion of a previous study~\cite{Xavier:2024iwr}.

\begin{figure}[H]
		\begin{center}
			\subfigure{\includegraphics[width=0.38\textwidth]{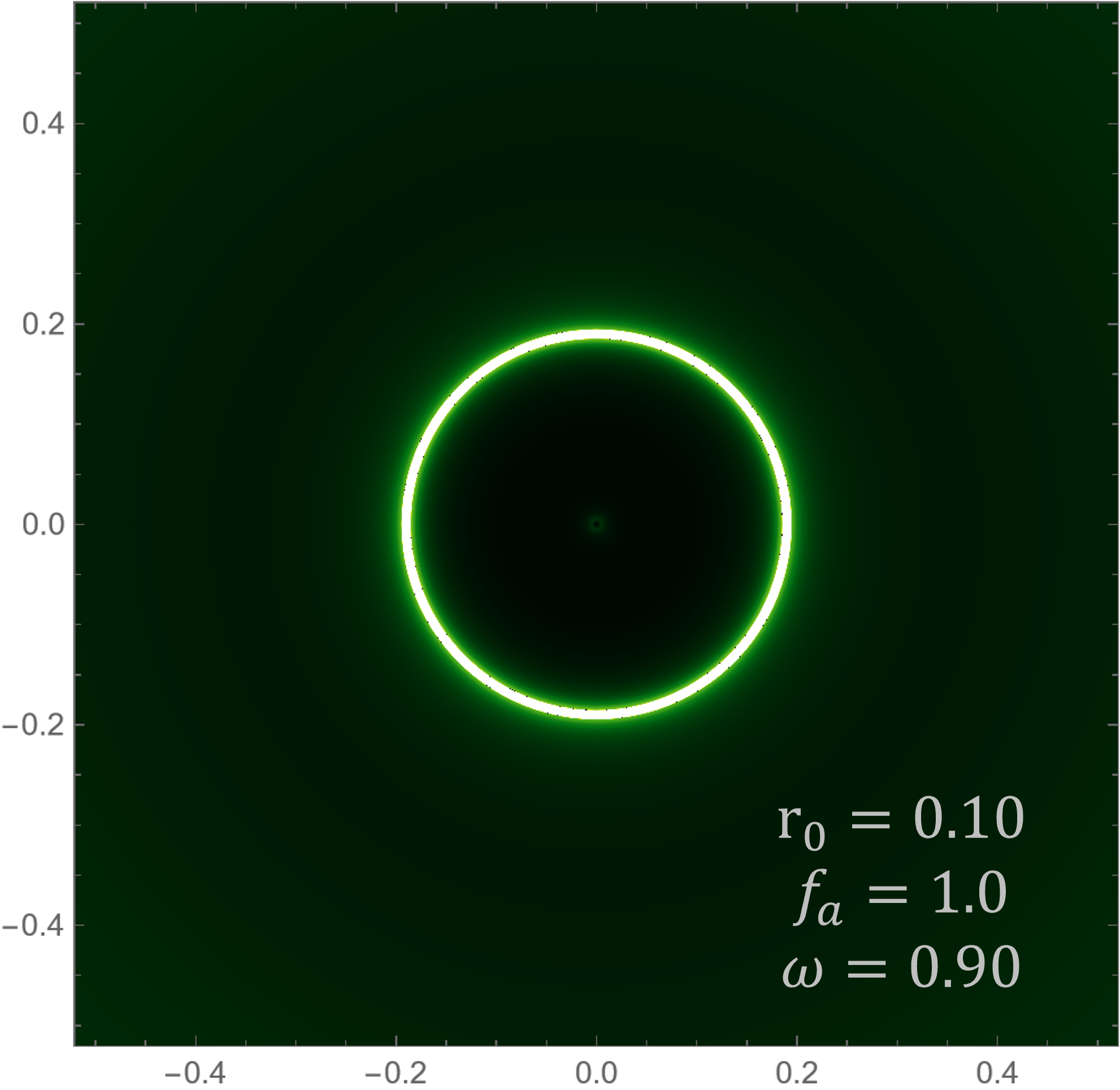}}
			\subfigure{\includegraphics[width=0.38\textwidth]{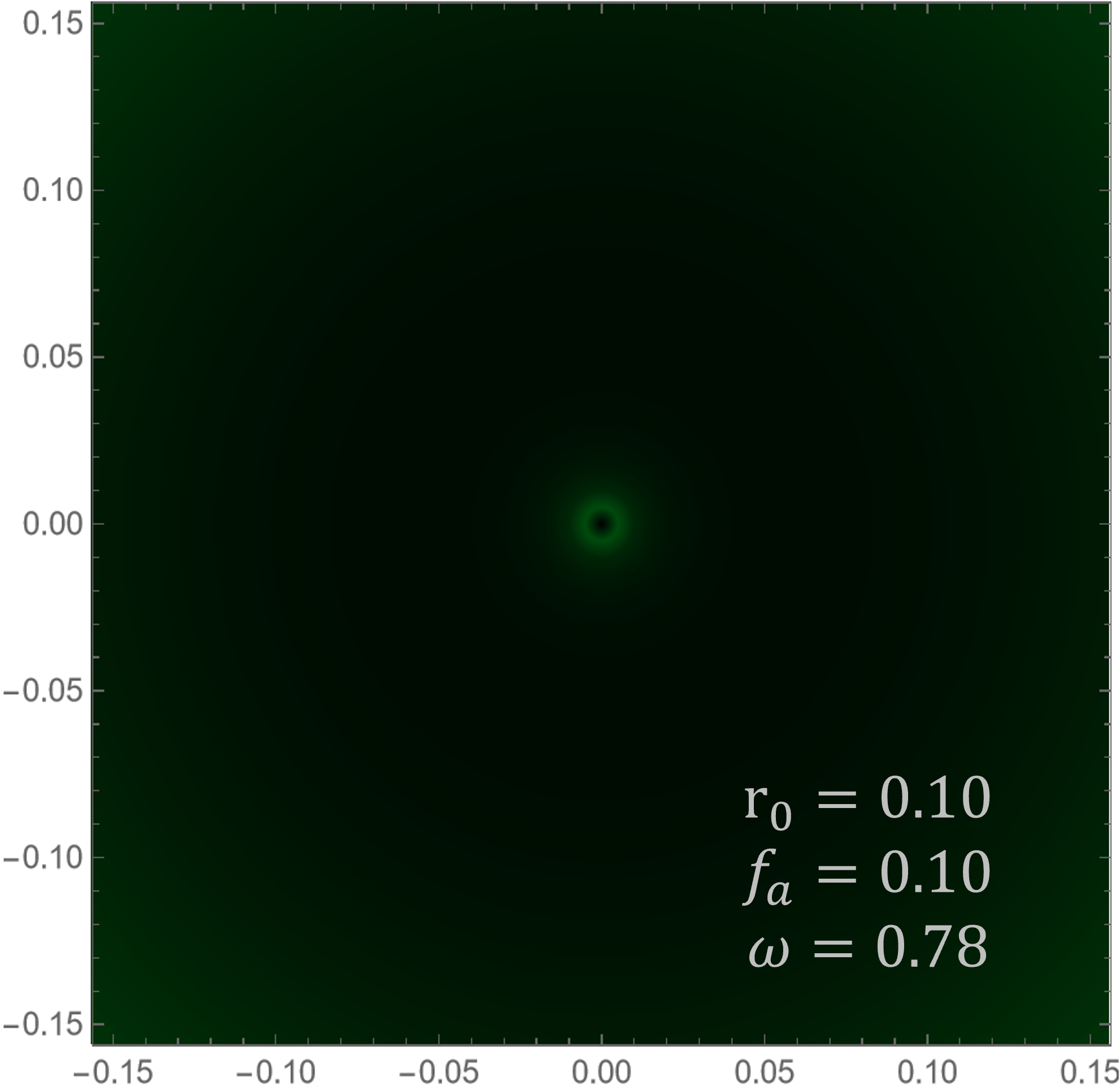}}
			\subfigure{\includegraphics[width=0.38\textwidth]{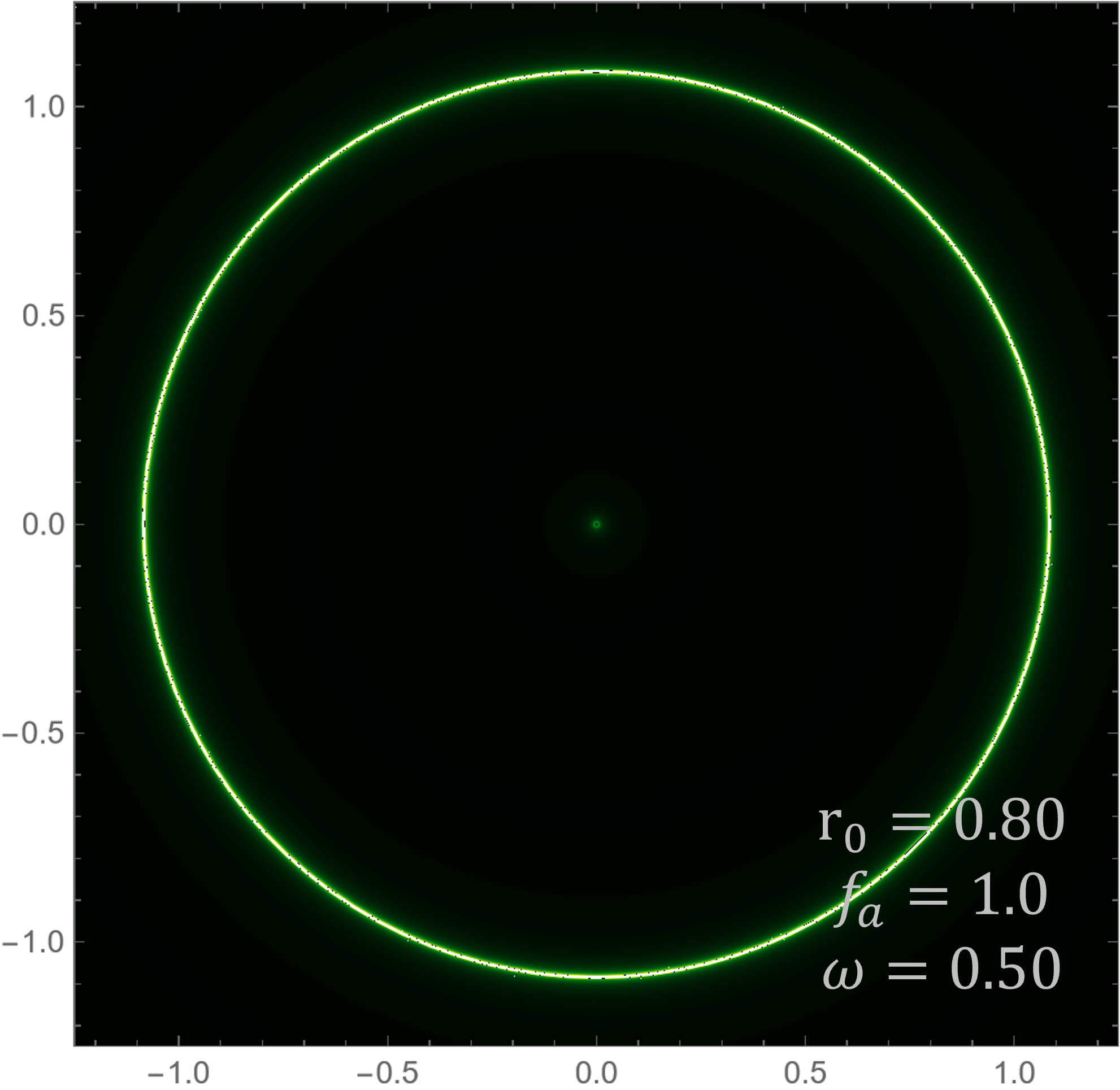}}
			\subfigure{\includegraphics[width=0.38\textwidth]{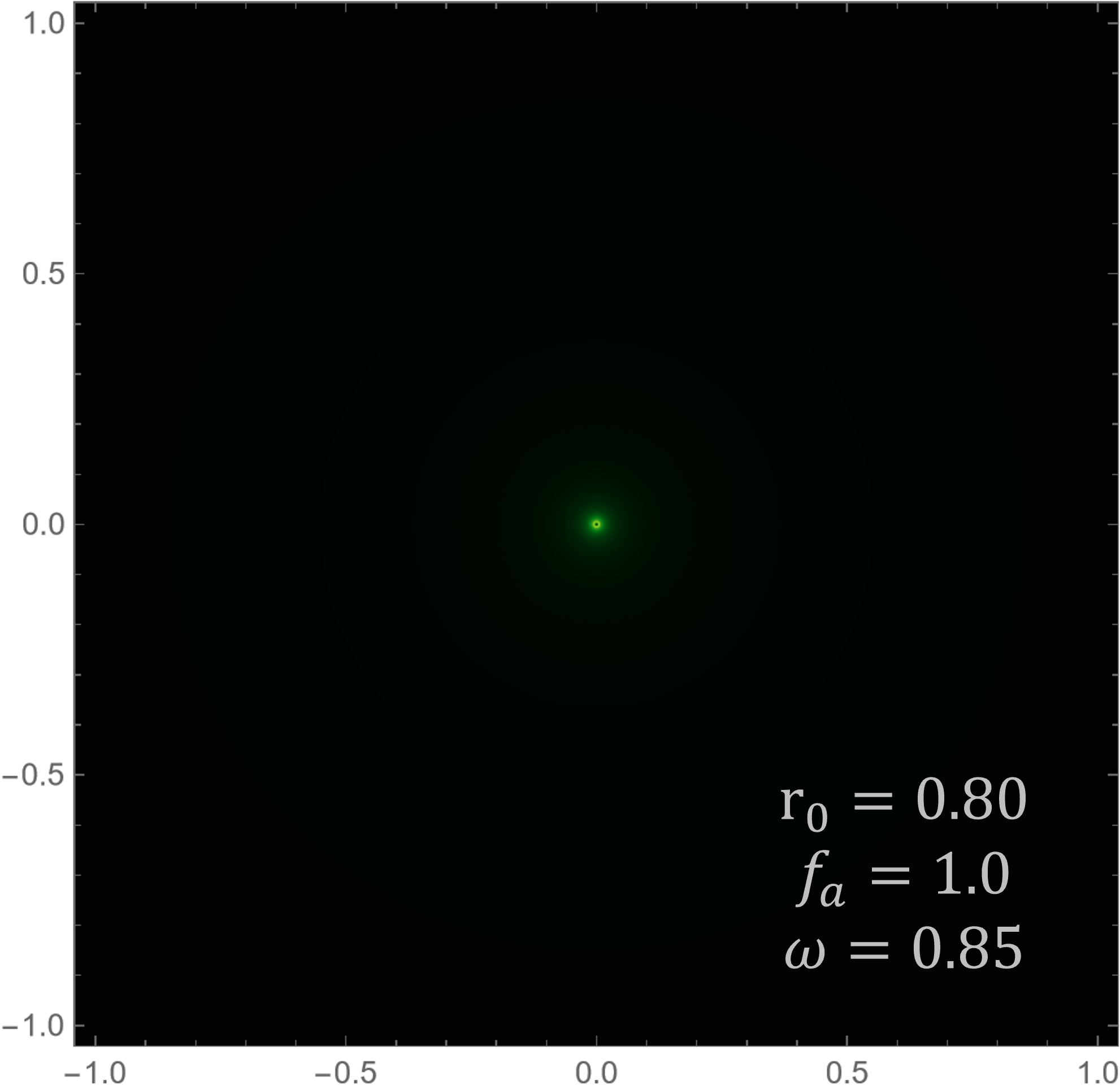}}
		\end{center}
		\caption{ The illustrations of the light ring for three different parameters. The left panel shows the solution with three light rings, while the right panel depicts a single photon ring at the throat.}
		\label{phase7}
	\end{figure}

	Moreover, Fig.~\ref{phase7} provides an illustrative diagram as for the light ring. We select the parameter sets ($r_0=0.1, f_a=1, \omega=0.9$) and ($r_0=0.8, f_a=1, \omega=0.5$) to illustrate the multi-LRs cases, and ($r_0=0.8, f_a=1, \omega=0.85$) and ($r_0=0.1, f_a=0.1, \omega=0.78$) to illustrate the single-LR cases.

	\section{Conclusion}\label{sec5}
	
	In this paper, we present a systematic investigation of a gravitational model featuring nontrivial spacetime topology, comprising an axion field as a dark matter candidate and a phantom field as a dark energy candidate. This model shares features with known phantom-energy wormhole solutions but also displays unique properties. As the $r_0$ increases and the $f_a$ decreases, the mass and charge curves versus frequency evolve from a tight spiral to a ``duckbill" shape, then to a loose spiral. For small $r_0$, the Kretschmann scalar is orders of magnitude larger and diverges near the throat, indicating a breakdown in traversability. Although it generally decreases with larger $r_0$, the scalar can increase at low frequencies-behavior linked to energy condition distributions. We find that, although the model typically violates the NEC near the throat, 
	in specific parameter regimes (e.g., for $r_0=0.5$ and low frequencies $\omega$), the NEC can be satisfied throughout the entire spacetime.  Furthermore, under other parameters (e.g., $r_0=0.8$),  even if the NEC remains violated at the throat,  symmetric regions where the NEC is satisfied emerge near the throat. Embedding diagrams show that the wormhole usually has one throat, but under special conditions, it can develop an equatorial plane and a double-throat structure.
	
	Furthermore, astronomical observational studies of wormhole models have been on the rise, including research on light rings and shadows \cite{Huang:2023yqd,Gjorgjieski:2025uik}, gravitational lensing \cite{Tsukamoto:2016qro}, and gravitational waves \cite{Cardoso:2016rao,DeSimone:2025sgu}. We analyze the null geodesics of our model as a simple yet powerful probe. It is found that the model either possesses a single unstable light ring at the throat, or, in addition, it has multiple pairs of light rings in the external spacetime. Regardless of the configuration, the number of unstable light rings is always one greater than the number of stable ones.
	Our analysis of traversability was limited to a preliminary study by computing the Kretschmann scalar at the throat of the static solution. However, a comprehensive understanding of a wormhole's traversability necessitates a full stability analysis \cite{Azad:2023iju} and the complete numerical evolution of a test particle as it traverses the wormhole \cite{Kain:2023ore}. These topics will constitute our next research direction.
	
	\section*{Acknowledgements}
	This work was supported by the National Natural Science Foundation of China under Grants  No. 12475051,  No. 12375051, and No. 12421005; the science and technology innovation Program of Hunan Province under grant No. 2024RC1050; the Natural Science Foundation of Hunan Province under grant No. 2023JJ30384; and  the innovative research group of Hunan Province under Grant No. 2024JJ1006.


\begin{thebibliography}{99}
		
		\bibitem{SupernovaSearchTeam:2001qse}
		A.~G.~Riess \textit{et al.} [Supernova Search Team],
		Astrophys. J. \textbf{560} (2001), 49-71
		doi:10.1086/322348
		[arXiv:astro-ph/0104455 [astro-ph]].
		
		\bibitem{Perlmutter:1999jt}
		S.~Perlmutter, M.~S.~Turner and M.~J.~White,
		Phys. Rev. Lett. \textbf{83} (1999), 670-673
		doi:10.1103/PhysRevLett.83.670
		[arXiv:astro-ph/9901052 [astro-ph]].
		
		\bibitem{WMAP:2003ivt}
		C.~L.~Bennett \textit{et al.} [WMAP],
		Astrophys. J. Suppl. \textbf{148} (2003), 1-27
		doi:10.1086/377253
		[arXiv:astro-ph/0302207 [astro-ph]].
		
		\bibitem{WMAP:2003zzr}
		G.~Hinshaw \textit{et al.} [WMAP],
		Astrophys. J. Suppl. \textbf{148} (2003), 135
		doi:10.1086/377225
		[arXiv:astro-ph/0302217 [astro-ph]].
		
		\bibitem{Cai:2004dk}
		R.~G.~Cai and A.~Wang,
		JCAP \textbf{03} (2005), 002
		doi:10.1088/1475-7516/2005/03/002
		[arXiv:hep-th/0411025 [hep-th]].
		
		\bibitem{Sahni:2002kh}
		V.~Sahni,
		Class. Quant. Grav. \textbf{19} (2002), 3435-3448
		doi:10.1088/0264-9381/19/13/304
		[arXiv:astro-ph/0202076 [astro-ph]].
		
		\bibitem{Copeland:2006wr}
		E.~J.~Copeland, M.~Sami and S.~Tsujikawa,
		Int. J. Mod. Phys. D \textbf{15} (2006), 1753-1936
		doi:10.1142/S021827180600942X
		[arXiv:hep-th/0603057 [hep-th]].
		
		\bibitem{Frieman:2008sn}
		J.~Frieman, M.~Turner and D.~Huterer,
		Ann. Rev. Astron. Astrophys. \textbf{46} (2008), 385-432
		doi:10.1146/annurev.astro.46.060407.145243
		[arXiv:0803.0982 [astro-ph]].
		
		\bibitem{Melchiorri:2002ux}
		A.~Melchiorri, L.~Mersini-Houghton, C.~J.~Odman and M.~Trodden,
		Phys. Rev. D \textbf{68} (2003), 043509
		doi:10.1103/PhysRevD.68.043509
		[arXiv:astro-ph/0211522 [astro-ph]].
		
		\bibitem{Alcaniz:2003qy}
		J.~S.~Alcaniz,
		Phys. Rev. D \textbf{69} (2004), 083521
		doi:10.1103/PhysRevD.69.083521
		[arXiv:astro-ph/0312424 [astro-ph]].
		
		\bibitem{Carroll:2003st}
		S.~M.~Carroll, M.~Hoffman and M.~Trodden,
		Phys. Rev. D \textbf{68} (2003), 023509
		doi:10.1103/PhysRevD.68.023509
		[arXiv:astro-ph/0301273 [astro-ph]].
		
		\bibitem{Amendola:2004qb}
		L.~Amendola,
		Phys. Rev. Lett. \textbf{93} (2004), 181102
		doi:10.1103/PhysRevLett.93.181102
		[arXiv:hep-th/0409224 [hep-th]].
		
		\bibitem{Planck:2015fie}
		P.~A.~R.~Ade \textit{et al.} [Planck],
		Astron. Astrophys. \textbf{594} (2016), A13
		doi:10.1051/0004-6361/201525830
		[arXiv:1502.01589 [astro-ph.CO]].
		
		\bibitem{Ellis:1973yv}
		H.~G.~Ellis,
		J. Math. Phys. \textbf{14} (1973), 104-118
		doi:10.1063/1.1666161.
		
		\bibitem{Ellis:1979bh}
		H.~G.~Ellis,
		Gen. Rel. Grav. \textbf{10} (1979), 105-123
		doi:10.1007/BF00756794.
		
		\bibitem{Kodama:1978dw}
		T.~Kodama,
		Phys. Rev. D \textbf{18} (1978), 3529-3534
		doi:10.1103/PhysRevD.18.3529.
		
		\bibitem{Einstein:1935tc}
		A.~Einstein and N.~Rosen,
		Phys. Rev. \textbf{48} (1935), 73-77,
		doi:10.1103/PhysRev.48.73.
		
		\bibitem{Misner:1957mt}
		C.~W.~Misner and J.~A.~Wheeler,
		Annals Phys. \textbf{2} (1957), 525-603,
		doi:10.1016/0003-4916(57)90049-0.
		
		\bibitem{Visser:1995cc}
		M.~Visser,
		``Lorentzian wormholes: From Einstein to Hawking,''
		Woodbury, USA: AIP (1995) 412.
		
		\bibitem{Lobo:2005us}
		F.~S.~N.~Lobo,
		Phys. Rev. D \textbf{71} (2005), 084011
		doi:10.1103/PhysRevD.71.084011
		[arXiv:gr-qc/0502099 [gr-qc]].
		
		\bibitem{Lobo:2005yv}
		F.~S.~N.~Lobo,
		Phys. Rev. D \textbf{71} (2005), 124022
		doi:10.1103/PhysRevD.71.124022
		[arXiv:gr-qc/0506001 [gr-qc]].
		
		\bibitem{Yazadjiev:2017twg}
		S.~Yazadjiev,
		Phys. Rev. D \textbf{96} (2017) no.4, 044045
		doi:10.1103/PhysRevD.96.044045
		[arXiv:1707.03654 [gr-qc]].
		
		\bibitem{Morris:1988cz}
		M.~S.~Morris and K.~S.~Thorne,
		Am. J. Phys. \textbf{56} (1988), 395-412
		doi:10.1119/1.15620
		
		
		\bibitem{Bronnikov:2012ch}
		K.~A.~Bronnikov, R.~A.~Konoplya and A.~Zhidenko,
		Phys. Rev. D \textbf{86} (2012), 024028
		doi:10.1103/PhysRevD.86.024028
		[arXiv:1205.2224 [gr-qc]].
		
		\bibitem{Kleihaus:2014dla}
		B.~Kleihaus and J.~Kunz,
		Phys. Rev. D \textbf{90} (2014), 121503
		doi:10.1103/PhysRevD.90.121503
		[arXiv:1409.1503 [gr-qc]].
		
		\bibitem{Novikov:2009vn}
		D.~I.~Novikov, A.~G.~Doroshkevich, I.~D.~Novikov and A.~A.~Shatskiy,
		Astron. Rep. \textbf{53} (2009), 1079-1085
		doi:10.1134/S1063772909120014
		[arXiv:0911.4456 [gr-qc]].
		
		\bibitem{Bronnikov:2013coa}
		K.~A.~Bronnikov, L.~N.~Lipatova, I.~D.~Novikov and A.~A.~Shatskiy,
		Grav. Cosmol. \textbf{19} (2013), 269-274
		doi:10.1134/S0202289313040038
		[arXiv:1312.6929 [gr-qc]].
		
		\bibitem{Huang:2020qmn}
		H.~Huang, H.~L{\"u} and J.~Yang,
		Class. Quant. Grav. \textbf{39} (2022) no.18, 185009
		doi:10.1088/1361-6382/ac8266
		[arXiv:2010.00197 [gr-qc]].
		
		\bibitem{Bronnikov:2002rn}
		K.~A.~Bronnikov and S.~W.~Kim,
		Phys. Rev. D \textbf{67} (2003), 064027
		doi:10.1103/PhysRevD.67.064027
		[arXiv:gr-qc/0212112 [gr-qc]].
		
		\bibitem{Kanti:2011jz}
		P.~Kanti, B.~Kleihaus and J.~Kunz,
		Phys. Rev. Lett. \textbf{107} (2011), 271101
		doi:10.1103/PhysRevLett.107.271101
		[arXiv:1108.3003 [gr-qc]].
		
		\bibitem{Blazquez-Salcedo:2020nsa}
		J.~L.~Bl{\'a}zquez-Salcedo, X.~Y.~Chew, J.~Kunz and D.~H.~Yeom,
		Eur. Phys. J. C \textbf{81} (2021) no.9, 858
		doi:10.1140/epjc/s10052-021-09645-0
		[arXiv:2012.06213 [gr-qc]].
		
		\bibitem{DeFalco:2023twb}
		V.~De Falco and S.~Capozziello,
		Phys. Rev. D \textbf{108} (2023) no.10, 104030
		doi:10.1103/PhysRevD.108.104030
		[arXiv:2308.05440 [gr-qc]].
		
		\bibitem{DeFalco:2023kqy}
		V.~De Falco,
		Phys. Rev. D \textbf{108} (2023) no.2, 024051
		doi:10.1103/PhysRevD.108.024051
		[arXiv:2307.03151 [gr-qc]].
		
		\bibitem{KordZangeneh:2015dks}
		M.~Kord Zangeneh, F.~S.~N.~Lobo and M.~H.~Dehghani,
		Phys. Rev. D \textbf{92} (2015) no.12, 124049
		doi:10.1103/PhysRevD.92.124049
		[arXiv:1510.07089 [gr-qc]].
		
		\bibitem{Harko:2013yb}
		T.~Harko, F.~S.~N.~Lobo, M.~K.~Mak and S.~V.~Sushkov,
		Phys. Rev. D \textbf{87} (2013) no.6, 067504
		doi:10.1103/PhysRevD.87.067504
		[arXiv:1301.6878 [gr-qc]].
		
		\bibitem{Konoplya:2021hsm}
		R.~A.~Konoplya and A.~Zhidenko,
		Phys. Rev. Lett. \textbf{128} (2022) no.9, 091104
		doi:10.1103/PhysRevLett.128.091104
		[arXiv:2106.05034 [gr-qc]].
		
		\bibitem{Blazquez-Salcedo:2020czn}
		J.~L.~Bl{\'a}zquez-Salcedo, C.~Knoll and E.~Radu,
		Phys. Rev. Lett. \textbf{126} (2021) no.10, 101102
		doi:10.1103/PhysRevLett.126.101102
		[arXiv:2010.07317 [gr-qc]].
		
		\bibitem{Kain:2023pvp}
		B.~Kain,
		Phys. Rev. D \textbf{108} (2023) no.8, 084010
		doi:10.1103/PhysRevD.108.084010
		[arXiv:2308.00049 [gr-qc]].
		
		\bibitem{Hu:2000ke}
		W.~Hu, R.~Barkana and A.~Gruzinov,
		Phys. Rev. Lett. \textbf{85} (2000), 1158-1161
		doi:10.1103/PhysRevLett.85.1158
		[arXiv:astro-ph/0003365 [astro-ph]].
		
		\bibitem{Sahni:1999qe}
		V.~Sahni and L.~M.~Wang,
		Phys. Rev. D \textbf{62} (2000), 103517
		doi:10.1103/PhysRevD.62.103517
		[arXiv:astro-ph/9910097 [astro-ph]].
		
		\bibitem{Matos:2000ng}
		T.~Matos and L.~A.~Urena-Lopez,
		Class. Quant. Grav. \textbf{17} (2000), L75-L81
		doi:10.1088/0264-9381/17/13/101
		[arXiv:astro-ph/0004332 [astro-ph]].
		
		\bibitem{Kaup:1968zz}
		D.~J.~Kaup,
		Phys. Rev. \textbf{172} (1968), 1331-1342
		doi:10.1103/PhysRev.172.1331
		
		\bibitem{Ruffini:1969qy}
		R.~Ruffini and S.~Bonazzola,
		Phys. Rev. \textbf{187} (1969), 1767-1783
		doi:10.1103/PhysRev.187.1767
		
		\bibitem{Schunck:1996he}
		F.~E.~Schunck and E.~W.~Mielke,
		Phys. Lett. A \textbf{249} (1998), 389-394
		doi:10.1016/S0375-9601(98)00778-6.
		
		\bibitem{Liebling:2012fv}
		S.~L.~Liebling and C.~Palenzuela,
		Living Rev. Rel. \textbf{26} (2023) no.1, 1
		doi:10.1007/s41114-023-00043-4
		[arXiv:1202.5809 [gr-qc]].
		
		
		\bibitem{Davidson:2016uok}
		S.~Davidson and T.~Schwetz,
		Phys. Rev. D \textbf{93} (2016) no.12, 123509
		doi:10.1103/PhysRevD.93.123509
		[arXiv:1603.04249 [astro-ph.CO]].
		
		\bibitem{Baer:2014eja}
		H.~Baer, K.~Y.~Choi, J.~E.~Kim and L.~Roszkowski,
		Phys. Rept. \textbf{555} (2015), 1-60
		doi:10.1016/j.physrep.2014.10.002
		[arXiv:1407.0017 [hep-ph]].
		
		\bibitem{Klaer:2017ond}
		V.~B.~Klaer and G.~D.~Moore,
		JCAP \textbf{11} (2017), 049
		doi:10.1088/1475-7516/2017/11/049
		[arXiv:1708.07521 [hep-ph]].
		
		\bibitem{Preskill:1982cy}
		J.~Preskill, M.~B.~Wise and F.~Wilczek,
		Phys. Lett. B \textbf{120} (1983), 127-132
		doi:10.1016/0370-2693(83)90637-8.
		
		\bibitem{Jaeckel:2010ni}
		J.~Jaeckel and A.~Ringwald,
		Ann. Rev. Nucl. Part. Sci. \textbf{60} (2010), 405-437
		doi:10.1146/annurev.nucl.012809.104433
		[arXiv:1002.0329 [hep-ph]].
		
		\bibitem{Arvanitaki:2010sy}
		A.~Arvanitaki and S.~Dubovsky,
		Phys. Rev. D \textbf{83} (2011), 044026
		doi:10.1103/PhysRevD.83.044026
		[arXiv:1004.3558 [hep-th]].
		
		\bibitem{Berezhiani:1992rk}
		Z.~G.~Berezhiani, A.~S.~Sakharov and M.~Y.~Khlopov,
		Sov. J. Nucl. Phys. \textbf{55} (1992), 1063-1071
		
		\bibitem{Sakharov:1996xg}
		A.~S.~Sakharov, D.~D.~Sokoloff and M.~Y.~Khlopov,
		Phys. Atom. Nucl. \textbf{59} (1996), 1005-1010.
		
		\bibitem{Khlopov:1999tm}
		M.~Y.~Khlopov, A.~S.~Sakharov and D.~D.~Sokoloff,
		Nucl. Phys. B Proc. Suppl. \textbf{72} (1999), 105-109
		doi:10.1016/S0920-5632(98)00511-8.
		
		\bibitem{Visinelli:2017ooc}
		L.~Visinelli, S.~Baum, J.~Redondo, K.~Freese and F.~Wilczek,
		Phys. Lett. B \textbf{777} (2018), 64-72
		doi:10.1016/j.physletb.2017.12.010
		[arXiv:1710.08910 [astro-ph.CO]].
		
		\bibitem{Braaten:2015eeu}
		E.~Braaten, A.~Mohapatra and H.~Zhang,
		Phys. Rev. Lett. \textbf{117} (2016) no.12, 121801
		doi:10.1103/PhysRevLett.117.121801
		[arXiv:1512.00108 [hep-ph]].
		
		\bibitem{Eby:2016cnq}
		J.~Eby, M.~Leembruggen, P.~Suranyi and L.~C.~R.~Wijewardhana,
		JHEP \textbf{12} (2016), 066
		doi:10.1007/JHEP12(2016)066
		[arXiv:1608.06911 [astro-ph.CO]].
		
		\bibitem{Guerra:2019srj}
		D.~Guerra, C.~F.~B.~Macedo and P.~Pani,
		JCAP \textbf{09} (2019) no.09, 061
		[erratum: JCAP \textbf{06} (2020) no.06, E01]
		doi:10.1088/1475-7516/2019/09/061
		[arXiv:1909.05515 [gr-qc]].
		
		\bibitem{Delgado:2020udb}
		J.~F.~M.~Delgado, C.~A.~R.~Herdeiro and E.~Radu,
		JCAP \textbf{06} (2020), 037
		doi:10.1088/1475-7516/2020/06/037
		[arXiv:2005.05982 [gr-qc]].
		
		\bibitem{Zeng:2021oez}
		Y.~B.~Zeng, S.~Y.~Cui, H.~B.~Li, S.~X.~Sun, Y.~P.~Zhang and Y.~Q.~Wang,
		Eur. Phys. J. C \textbf{84} (2024) no.2, 187
		doi:10.1140/epjc/s10052-024-12536-9
		[arXiv:2103.10717 [gr-qc]].
		
		\bibitem{Chen:2023vet}
		J.~R.~Chen, L.~X.~Huang, L.~Zhao and Y.~Q.~Wang,
		Phys. Rev. D \textbf{109} (2024) no.10, 104078
		doi:10.1103/PhysRevD.109.104078
		[arXiv:2311.11830 [gr-qc]].
		
		\bibitem{Dzhunushaliev:2014bya}
		V.~Dzhunushaliev, V.~Folomeev, C.~Hoffmann, B.~Kleihaus and J.~Kunz,
		Phys. Rev. D \textbf{90} (2014) no.12, 124038
		doi:10.1103/PhysRevD.90.124038
		[arXiv:1409.6978 [gr-qc]].
		
		\bibitem{Ding:2023syj}
		P.~B.~Ding, T.~X.~Ma, T.~F.~Fang and Y.~Q.~Wang,
		JHEP \textbf{04} (2024), 033
		doi:10.1007/JHEP04(2024)033
		[arXiv:2305.19819 [gr-qc]].
		
		
		\bibitem{Hao:2024hba}
		C.~H.~Hao, X.~Su and Y.~Q.~Wang,
		Eur. Phys. J. C \textbf{85} (2025) no.3, 348
		doi:10.1140/epjc/s10052-025-13937-0
		[arXiv:2404.11002 [gr-qc]].
		
		\bibitem{Su:2023zhh}
		X.~Su, C.~H.~Hao, J.~R.~Ren and Y.~Q.~Wang,
		JCAP \textbf{09} (2024), 010
		doi:10.1088/1475-7516/2024/09/010
		[arXiv:2311.17557 [gr-qc]].
		
		\bibitem{Hao:2023igi}
		C.~H.~Hao, S.~X.~Sun, L.~X.~Huang, R.~Zhang, X.~Su and Y.~Q.~Wang,
		JCAP \textbf{04} (2024), 057
		doi:10.1088/1475-7516/2024/04/057
		[arXiv:2309.16379 [gr-qc]].
		
		\bibitem{Su:2024gxp}
		X.~Su, C.~H.~Hao and Y.~Q.~Wang,
		[arXiv:2407.09591 [gr-qc]].
		
		\bibitem{Hoffmann:2017jfs}
		C.~Hoffmann, T.~Ioannidou, S.~Kahlen, B.~Kleihaus and J.~Kunz,
		Phys. Rev. D \textbf{95} (2017) no.8, 084010
		doi:10.1103/PhysRevD.95.084010
		[arXiv:1703.03344 [gr-qc]].
		
		\bibitem{Dzhunushaliev:2025fbf}
		V.~Dzhunushaliev, V.~Folomeev and S.~Sakhiyev,
		Gen. Rel. Grav. \textbf{57} (2025) no.8, 126
		doi:10.1007/s10714-025-03463-5
		[arXiv:2505.05818 [gr-qc]].
		
		\bibitem{Hao:2023kvf}
		C.~H.~Hao, L.~X.~Huang, X.~Su and Y.~Q.~Wang,
		Eur. Phys. J. C \textbf{84} (2024) no.8, 878
		doi:10.1140/epjc/s10052-024-13105-w
		[arXiv:2312.03800 [gr-qc]].
		
		\bibitem{Wang:2016lxa}
		B.~Wang, E.~Abdalla, F.~Atrio-Barandela and D.~Pavon,
		Rept. Prog. Phys. \textbf{79} (2016) no.9, 096901
		doi:10.1088/0034-4885/79/9/096901
		[arXiv:1603.08299 [astro-ph.CO]].
		
		\bibitem{GrillidiCortona:2015jxo}
		G.~Grilli di Cortona, E.~Hardy, J.~Pardo Vega and G.~Villadoro,
		JHEP \textbf{01} (2016), 034
		doi:10.1007/JHEP01(2016)034
		[arXiv:1511.02867 [hep-ph]].
		
		\bibitem{Cunha:2022gde}
		P.~Cunha, V.P., C.~Herdeiro, E.~Radu and N.~Sanchis-Gual,
		Phys. Rev. Lett. \textbf{130} (2023) no.6, 061401
		doi:10.1103/PhysRevLett.130.061401
		[arXiv:2207.13713 [gr-qc]].
		
		\bibitem{Xavier:2024iwr}
		S.~V.~M.~C.~B.~Xavier, C.~A.~R.~Herdeiro and L.~C.~B.~Crispino,
		Phys. Rev. D \textbf{109} (2024) no.12, 124065
		doi:10.1103/PhysRevD.109.124065
		[arXiv:2404.02208 [gr-qc]].
		
		\bibitem{Kain:2023ore}
		B.~Kain,
		Phys. Rev. Lett. \textbf{131} (2023) no.10, 101001
		doi:10.1103/PhysRevLett.131.101001
		[arXiv:2309.03314 [hep-th]].
		
		\bibitem{Huang:2023yqd}
		H.~Huang, J.~Kunz, J.~Yang and C.~Zhang,
		Phys. Rev. D \textbf{107} (2023) no.10, 104060
		doi:10.1103/PhysRevD.107.104060
		[arXiv:2303.11885 [gr-qc]].
		
		\bibitem{Gjorgjieski:2025uik}
		K.~Gjorgjieski, J.~Kunz and P.~Nedkova,
		[arXiv:2505.07507 [gr-qc]].
		
		\bibitem{Tsukamoto:2016qro}
		N.~Tsukamoto,
		Phys. Rev. D \textbf{94} (2016) no.12, 124001
		doi:10.1103/PhysRevD.94.124001
		[arXiv:1607.07022 [gr-qc]].
		
		\bibitem{Cardoso:2016rao}
		V.~Cardoso, E.~Franzin and P.~Pani,
		Phys. Rev. Lett. \textbf{116} (2016) no.17, 171101
		[erratum: Phys. Rev. Lett. \textbf{117} (2016) no.8, 089902]
		doi:10.1103/PhysRevLett.116.171101
		[arXiv:1602.07309 [gr-qc]].
		
		\bibitem{DeSimone:2025sgu}
		C.~De Simone, V.~De Falco and S.~Capozziello,
		Phys. Rev. D \textbf{111} (2025) no.6, 064021
		doi:10.1103/PhysRevD.111.064021
		[arXiv:2502.12646 [gr-qc]].
		
		\bibitem{Azad:2023iju}
		B.~Azad, J.~L.~Blazquez-Salcedo, F.~S.~Khoo and J.~Kunz,
		Phys. Lett. B \textbf{848} (2024), 138349
		doi:10.1016/j.physletb.2023.138349
		[arXiv:2301.05243 [gr-qc]].
		
		\bibitem{Lee:1988av}
		T.~D.~Lee and Y.~Pang,
		Nucl. Phys. B \textbf{315} (1989), 477
		doi:10.1016/0550-3213(89)90365-9
		
		\bibitem{Gleiser:1988rq}
		M.~Gleiser,
		Phys. Rev. D \textbf{38} (1988), 2376
		[erratum: Phys. Rev. D \textbf{39} (1989) no.4, 1257]
		doi:10.1103/PhysRevD.38.2376.
		
		\bibitem{Herdeiro:2021lwl}
		C.~A.~R.~Herdeiro, A.~M.~Pombo, E.~Radu, P.~V.~P.~Cunha and N.~Sanchis-Gual,
		JCAP \textbf{04} (2021), 051
		doi:10.1088/1475-7516/2021/04/051
		[arXiv:2102.01703 [gr-qc]].
		
		
	\end{thebibliography}
\end{document}